\documentclass[conference,compsoc]{IEEEtran}
\IEEEoverridecommandlockouts

\ifCLASSOPTIONcompsoc
  \usepackage[nocompress]{cite}
\else
  \usepackage{cite}
\fi
\ifCLASSINFOpdf
\else
\fi
\usepackage{tikz}

\usepackage[T1]{fontenc}
\usepackage[utf8]{inputenc}
\usepackage{xcolor}
\usepackage{xspace}
\usepackage{bm}
\usepackage{url}
\usepackage{verbatim}
\usepackage{hyperref}
\usepackage{setspace}

\usepackage{amsmath,amsfonts}
\DeclareMathOperator*{\argmin}{argmin}
\usepackage{amssymb}

\usepackage{caption}
\usepackage{booktabs}
\usepackage{multirow}
\usepackage{makecell}

\usepackage{graphicx}
\usepackage{stfloats}
\usepackage{subfig}

\usepackage[ruled,vlined]{algorithm2e}
\usepackage{algorithmic}
\usepackage{listings}

\usepackage{cite}

\usepackage{enumitem}
\usepackage{textcomp}

\usepackage{array}

\def\red#1{\textcolor{red}{#1}}
\def\orange#1{\textcolor{orange}{#1}}
\newcommand{\ie}{\textit{i.e.}}
\newcommand{\eg}{\textit{e.g.}}

\newcommand{\partitle}[1]{\medskip \noindent \textbf{#1.}}
\newcommand{\Name}{\texttt{PromptCOS}\xspace}

\begin{document}
%
\title{\Large \bf \Name: Towards Content-only System Prompt Copyright Auditing for LLMs}

\author{
Yuchen Yang$^{1}$,
Yiming Li$^{2}$\textsuperscript{*}\thanks{\textsuperscript{*}Corresponding Author: Yiming Li (liyiming.tech@gmail.com).},
Hongwei Yao$^{1}$,
Enhao Huang$^{1}$,
Shuo Shao$^{1}$,
Yuyi Wang$^{3}$,\\
Zhibo Wang$^{1}$,
Dacheng Tao$^{2}$,
Zhan Qin$^{1}$ \\
$^{1}$State Key Laboratory of Blockchain and Data Security, Zhejiang University \\
$^{2}$College of Computing and Data Science, Nanyang Technological University \\
$^{3}$CRRC Zhuzhou Institute \& Tengen Intelligence Institute \\
\texttt{\small\{ychyang, huangenhao, shaoshuo\_ss, zhibowang, qinzhan\}@zju.edu.cn} \\
\texttt{\small \{liyiming.tech, dacheng.tao, yuyiwang920\}@gmail.com; yao.hongwei@cityu.edu.hk}
}


%


\maketitle

\begin{abstract}
System prompts are critical for shaping the behavior and output quality of large language model (LLM)–based applications, driving substantial investment in optimizing high-quality prompts beyond traditional handcrafted designs. However, as system prompts become valuable intellectual property, they are increasingly vulnerable to prompt theft and unauthorized use, highlighting the urgent need for effective copyright auditing, especially watermarking. Existing methods rely on verifying subtle logit distribution shifts triggered by a query. We observe that this logit-dependent verification framework is impractical in real-world content-only settings, primarily because (1) random sampling makes content-level generation unstable for verification, and (2) stronger instructions needed for content-level signals compromise prompt fidelity.

To overcome these challenges, we propose \Name, the first content-only system prompt copyright auditing method based on content-level output similarity. \Name achieves watermark stability by designing a cyclic output signal as the conditional instruction's target. It preserves prompt fidelity by injecting a small set of auxiliary tokens to encode the watermark, leaving the main prompt untouched. Furthermore, to ensure robustness against malicious removal, we optimize cover tokens, \ie, critical tokens in the original prompt, to ensure that removing auxiliary tokens causes severe performance degradation. Experimental results show that \Name achieves high effectiveness (99.3\% average watermark similarity), strong distinctiveness (60.8\% higher than the best baseline), high fidelity (accuracy degradation no greater than 0.6\%), robustness (resilience against four potential attack categories), and high computational efficiency (up to 98.1\% cost saving).

\end{abstract}


%
\IEEEpeerreviewmaketitle

\section{Introduction}
\label{sec:intro}

Large Language Models (LLMs) have recently achieved remarkable progress. Advanced models such as ChatGPT \cite{openai2024gpt4technicalreport} and DeepSeek \cite{deepseekai2025deepseekr1incentivizingreasoningcapability} demonstrate outstanding linguistic and reasoning capabilities, enabling the generation of high-quality content with significant commercial value \cite{openai2024gpt4technicalreport}. Consequently, developers have increasingly integrated LLMs into various applications to provide intelligent services, leading to the rapid emergence of LLM-based applications (\eg, chatbots \cite{ChatGPT} and AI consultants \cite{Quick-Suite}).


A critical factor in developing LLM-based applications is the \emph{system prompt}, which directly influences the model's behavior and output quality \cite{hou2024large, yang2025prsa}. Indeed, for application developers leveraging large-scale pre-trained models, where modifying parameters is often computationally expensive or even infeasible, the system prompt serves as a primary mechanism for steering model alignment and task-specific adaptation. Initial system prompts were manually-designed (\eg, ``You are a helpful AI assistant''). However, the manual design is highly dependent on human experts, often yields suboptimal performance, and the variety of potential handcrafted prompts is limited \cite{yu2023exploringpromptengineering,nazzal2024promsec}. To overcome these limitations, recent efforts \cite{shin2020autopromptelicitingknowledgelanguage,zhang2022automaticchainthoughtprompting} focus on automated prompt generation to derive specialized system prompts. These prompts are often model-specific and non-human-readable, designed to maximize task efficacy. Beyond their substantial commercial value, their creation requires considerable computational resources, expert knowledge, and proprietary training data. Consequently, system prompts are emerging as a novel form of intellectual property in the generative AI era~\cite{li2025rethinking}, giving rise to specialized prompt markets~\cite{promptmarket2025promptbase} and warranting dedicated protection.

\begin{figure}[t]
  \centering %
  \includegraphics[width=0.45\textwidth]{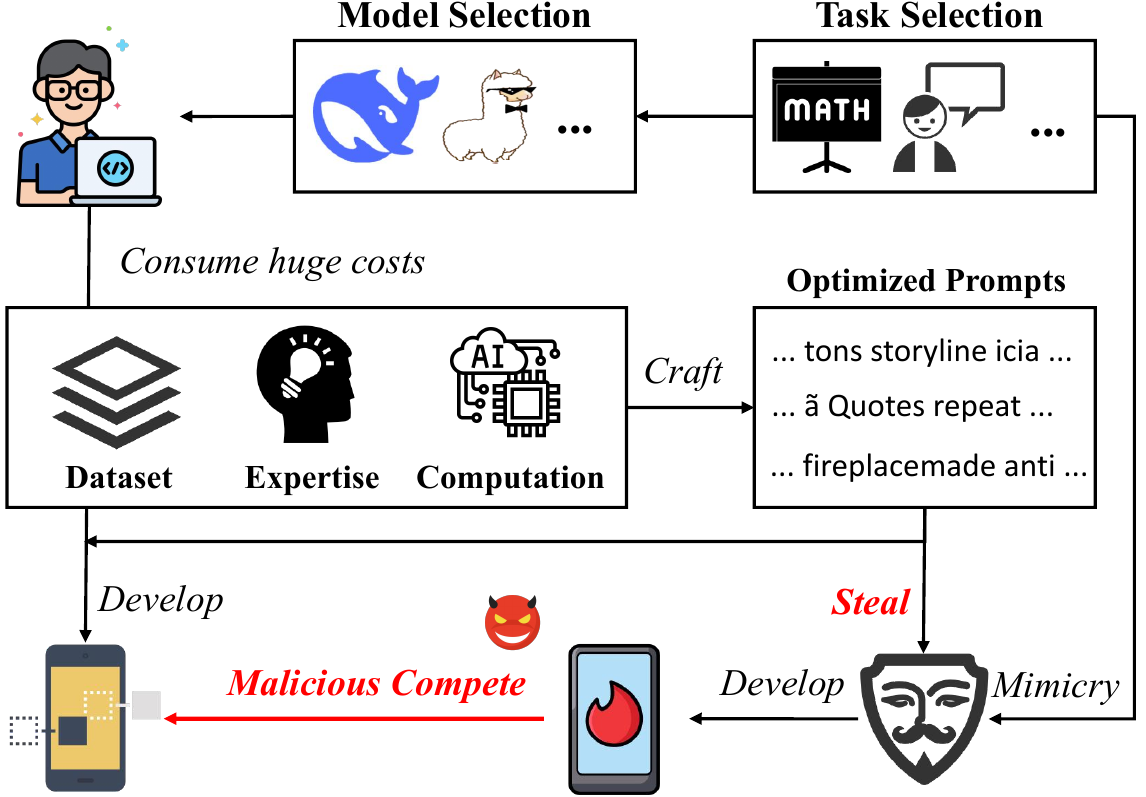} 
  \caption{Illustration of prompt leakage. To develop an LLM-based application, the prompt owner designs a high-quality system prompt requiring significant computational resources and expert data. However, adversaries may steal the prompt through illegal prompt deals \cite{promptmarket2025promptbase} or data leakage \cite{agarwal2024promptleakage,yang2025prsa} to develop competitive applications, harming the interests of the prompt owner.} 
  \label{fig:prompt_leakage} 
  \vspace{-1em}
\end{figure}

Unfortunately, system prompts are increasingly exposed to infringement risks, giving rise to serious copyright and security concerns. Recent studies have demonstrated that system prompts are highly vulnerable to theft and leakage attacks~\cite{PLeakccs24,agarwal2024promptleakage,yang2025prsa}. Alarmingly, such prompt leakage incidents have been observed in real-world deployments, affecting major LLM-based applications including Microsoft Bing~\cite{PromptLeakage2025MicrosoftBing} and GPT-5~\cite{promptleakage2025GPT5}. As illustrated in Figure~\ref{fig:prompt_leakage}, malicious developers can replicate or even commercialize competing applications by stealing existing system prompts, without requiring critical development resources. Recognizing its severity, the OWASP Foundation has listed prompt leakage among the top ten security risks for LLMs~\cite{LLMtop10}.

Conventional data protection techniques, such as data encryption~\cite{Posterccs24encryption} and access control~\cite{yang2024pmaccess}, are inherently inadequate for safeguarding system prompts. This inadequacy stems from the operational nature of LLM-based applications: to guide model behavior, the system prompt must be transmitted to the model during every inference. Consequently, the LLM itself becomes an intrinsic channel through which the system prompt can be accessed, either directly or indirectly, by adversaries. To address this challenge, watermark-based copyright auditing methods are recently proposed to protect the system prompt \cite{yao2024promptcare}. These methods modify the original system prompt by embedding a \emph{conditional instruction}, \ie, optimizing the prompt token by token to interact with an owner-defined \emph{trigger}. The presence of the trigger activates the instruction, inducing distinctive generation behaviors. Under normal queries, LLM-based applications using the watermarked prompt behave similarly to those using the original prompt. However, when queries contain the trigger, the watermarked prompt enforces copyright-related responses (\eg, exhibiting a distinct logit distribution). In particular, depending on the application scenarios of system prompts, a valid prompt watermarking scheme should at least satisfy four main requirements: \textbf{(1)} effectiveness, \textbf{(2)} distinctiveness, \textbf{(3)} fidelity, and \textbf{(4)} robustness. Specifically, effectiveness requires that the owner can obtain expected information if their system prompt is stolen; distinctiveness ensures the owner can clearly differentiate between watermarked and normal prompts, thereby minimizing false positives; fidelity ensures that embedding the watermark does not significantly degrade the utility of the prompt; and robustness ensures resilience against adversarial removal or evasion. To the best of our knowledge, only injection-based prompt watermarking methods \cite{li2023protecting, yao2024promptcare} simultaneously satisfy all four requirements for system prompt copyright protection. In general, after constructing a training dataset and a verification dataset to optimize the prompt, this type of method influences the logit distribution for generating the next token of the \emph{trigger} and uses hypothesis testing for copyright verification.

In this paper, we observe that these methods rely on an implicit assumption: the defender (\ie, the prompt owner) has access to the intermediate results (\eg, logits) of the suspicious third-party LLM-based application. However, this assumption usually does not hold in practice \cite{he2024difficulty}, where the defender can only get the generated content (without intermediate results) of the query from the suspicious applications. 
It is reasonable for developers including adversaries to refuse to share intermediate results. Specifically, mainstream application scenarios such as chatbots \cite{si2022so,li2024citation} and search tools \cite{sharma2024generative,gong2024cosearchagent} typically do not expose logits interfaces; besides, sharing intermediate outputs may introduce security risks, such as knowledge stealing \cite{panda2024teach} or data leakage \cite{li2024governing}. In particular, we reveal that simply extending existing logits-available auditing methods to the content-only setting is not practical to a large extent. 
Consequently, an intriguing question arises: \emph{can we design an effective content-only auditing method for protecting the copyright of system prompts?}

Fortunately, the answer to the above important question is positive, though achieving it is non-trivial. Designing an effective content-only prompt auditing method faces two main challenges. Firstly, unlike logit-level outputs, content-level outputs are inherently \emph{unpredictable} due to the random sampling mechanism, making the existing verification framework based on the next token of trigger incapable of guaranteeing stable watermark generation. Secondly, stronger conditional instructions require further modifications to the original prompt, which may lead to a decrease in fidelity. To overcome these challenges, we propose a simple yet effective prompt copyright auditing method, dubbed \Name, based on content-level output similarity. \Name designs conditional instructions that induce cyclic outputs, thereby stabilizing watermark generation (\ie, Challenge 1). Furthermore, it alleviates the fidelity–instruction conflict (\ie, Challenge 2) by introducing auxiliary tokens to preserve the prompt's fidelity while maintaining auditing effectiveness.

Specifically, \Name consists of two main phases: \emph{watermark embedding} and \emph{copyright verification}. In the \emph{watermark embedding} phase, we jointly optimize the system prompt, the verification query, and the signal mark (\ie, the desired verification output). This process involves three key designs:
\textbf{(1)} \emph{Cyclic Output Signals for Effectiveness}: we adopt a cyclic content structure containing the signal mark as the target output of the conditional instruction, which ensures more stable signal generation.
\textbf{(2)} \emph{Auxiliary Tokens for Fidelity}: we inject and modify only a small set of auxiliary tokens to encode semantic information, leaving the main system prompt unchanged to preserve fidelity. 
\textbf{(3)} \emph{Cover Tokens for Robustness}: to resist malicious deletion of auxiliary tokens, we optimize a small number of cover tokens (\ie, critical tokens in the original prompt) such that removing auxiliary tokens degrades the performance of the watermarked prompt. 
To achieve these objectives, we employ an alternating optimization algorithm with a greedy gradient search to refine the prompt, verification query, and signal mark at the token level. In the \emph{copyright verification} phase, we submit the verification query to the suspicious LLM-based application and assess its outputs. Specifically, we segment the generated content using a sliding window and compute its similarity with the signal mark. If the maximum similarity score exceeds a predefined threshold, the system prompt is deemed misappropriated.

We conduct extensive experiments to evaluate our \Name. Specifically, we exploit three popular LLM series (Llama \cite{touvron2023llamaopenefficientfoundation}, Gemma \cite{team2024gemma} and Deepseek \cite{deepseekai2025deepseekr1incentivizingreasoningcapability}), and three representative benchmark datasets. Our experiments cover all mainstream tasks used for benchmarking LLMs \cite{openai2024gpt4technicalreport,deepseekai2025deepseekr1incentivizingreasoningcapability}, including question-answering, mathematical reasoning, and programing. 
The results demonstrate that our \Name achieves effectiveness (99.3\% average watermark similarity), distinctiveness (60.8\% greater than the best baseline), fidelity (accuracy degradation of no more than 0.58\%), robustness (resistance to four types of potential attacks), and high computational efficiency (up to 98.1\% computational cost savings). 

Our main contributions are three-fold: \textbf{1)} We analyze the mechanisms of existing prompt copyright auditing methods and show that their reliance on intermediate model outputs limits their applicability in more challenging and realistic content-only scenarios. \textbf{2)} We propose \Name, the first content-only prompt auditing method with effectiveness, distinctiveness, fidelity, robustness and higher efficiency.  \textbf{3)} We conduct extensive experiments to verify the superior performance of our \Name and its resistance to potential adaptive attacks.

\section{Background and Related Work}

\subsection{LLMs and LLM-based Applications}

Large Language Models (LLMs) represent a significant advancement in natural language processing. These models are primarily transformer-based neural networks, trained on vast amounts of text data, and exhibit strong capabilities in various domains, such as translation, question answering, and code generation~\cite{openai2024gpt4technicalreport, he2025benchmarking}. Currently, most state-of-the-art LLMs are autoregressive, predicting the probability distribution for the next token based on preceding tokens \cite{openai2024gpt4technicalreport,deepseekai2025deepseekr1incentivizingreasoningcapability}. Given a user input, an LLM can generate a response by iteratively predicting the next token.

Based on the strong capabilities of large language models (LLMs), developers can build applications that support complex generative tasks (\eg, fluent dialogue \cite{ChatGPT}, code generation \cite{github_copilot}), collectively referred to as LLM-based applications. Since the effectiveness of such applications is predominantly determined by the input–output behaviors of the underlying LLM rather than the surrounding development pipeline \cite{hou2025securityLLMapplication}, improving model performance on target tasks becomes essential for enhancing overall application quality \cite{lin2024parrotLLMapplication}. A direct approach is to update the LLM through fine-tuning or other personalization techniques. However, such parameter-level modification requires substantial computational resources and may introduce undesirable effects such as catastrophic forgetting \cite{huang2024catastrophic_forgetting}, as well as additional security risks \cite{yi2025probebackdoor,sun2025peftguard}. An alternative approach leverages the strong in-context learning capabilities of LLMs by carefully optimizing the preset textual instructions (\ie, system prompts) used in the application \cite{nazzal2024promsec}. Compared to parameter-level adaptation, this prompt-based strategy is more efficient and has therefore been widely adopted in practice.

\subsection{System Prompts and Prompt Engineering}
\label{sec:system prompts and prompt engineering}

Given the critical role of system prompts, designing high-quality prompts has emerged as a central research problem. The practice of crafting effective prompts, known as prompt engineering, has evolved substantially to better exploit the capabilities of LLMs. Early efforts primarily relied on manually constructed system prompts derived from practical intuition, such as adding supplementary instructions (``You are a helpful assistant'') or specializing the generation process (``Let us think step by step'') \cite{wei2022chainCoT}. Although these approaches are easy to implement, they offer only coarse-grained control over model behavior and lack mechanisms to incorporate model-driven feedback during prompt refinement. In addition, intuition-based prompt designs are inherently limited in types, and their effectiveness often fails to generalize across tasks or meet developer expectations \cite{yu2023exploringpromptengineering,nazzal2024promsec}. These limitations substantially restrict their ability to enhance model performance on complex tasks.

To address these limitations, recent research has focused on automating the optimization of system prompts to maximize LLM performance on target tasks. These methods typically start from an initial prompt and iteratively refine it through either token-level optimization \cite{shin2020autopromptelicitingknowledgelanguage} or model-based prompt generation techniques \cite{zhou2022large}. The optimization process is guided by feedback obtained from evaluation benchmarks (\ie, task-specific training datasets) in conjunction with the target model. We formalize this general optimization paradigm in Section~\ref{sec:formulation of prompt engineering}. A notable characteristic of optimization-based prompt engineering is that the resulting prompts are often non-human-readable, as illustrated in Figure~\ref{fig:prompt_leakage}. However, since system prompts are typically concealed from end users in real-world applications \cite{shen2023promptcv}, the lack of readability does not diminish their practical utility. At the same time, the optimization pipeline incurs substantial costs, including the need for labeled data, domain expertise, and non-trivial computational resources \cite{yao2024promptcare}. These costs far exceed those associated with manual design, and the resulting performance gains translate directly into competitive advantages in downstream products, making optimized system prompts a valuable form of intellectual property for LLM-based applications.

\subsection{Watermarking Techniques for LLMs}
As LLMs become increasingly powerful and valuable, ensuring their traceability and ownership is becoming more and more critical. Watermarking has emerged as a promising solution to address these issues~\cite{liu2024survey}. Broadly, existing watermarks for LLMs can be classified into generative content (GC), model, and prompt watermarking~\cite{zhao2024watermarking, ren2024sok}.

\partitle{Generative Content (GC) and Model Watermarking} The majority of the existing watermarking methods focus on the former two categories~\cite{ren2024sok}. GC watermarking refers to embedding identifiable information (\ie, watermark) into the text output produced by an LLM~\cite{kirchenbauer2023watermark, kuditipudirobust, zhang2024remark}. The watermark can be embedded into GC either during or after the generation of texts~\cite{yang2022tracing, zhao2024provable}. GC watermarking can detect whether a paragraph of text is generated by an LLM. In contrast, model watermarking aims to embed watermarks into the models and protect their copyright~\cite{shao2025explanation}. In case the watermarked model is stolen by an adversary, the model developer can extract the watermark inside the model and accuse the adversary of infringement~\cite{krauss2024clearstamp, xu2024instructional}. The primary difference between these two watermarking methods is that GC watermarking extracts watermarks from GC directly, while model watermarking does so through interactions with the model (\ie, input and output pairs). However, both methods function by influencing the generated content, either through modifying model parameters or directly altering the output, regardless of whether their primary target is the model or the content itself. In contrast, system prompts are physically independent of the LLM, although they still affect the final generated content through their role within the LLM-integrated system. Consequently, these watermarking techniques arguably cannot be applied directly to audit the copyright of system prompts.

\partitle{Prompt Watermarking}
Existing prompt watermarking methodologies \cite{li2023protecting,yao2024promptcare} operate by embedding covert semantic signals within system prompts via token-level modifications. This strategic intervention induces the prompt to yield stealthy yet distinct outputs, such as special logit distributions, contingent upon the presence of a predefined \emph{trigger} within the user query. If the query does not contain the trigger, the generation process will not be interfered with. However, these approaches are predicated upon the accessibility of the LLM's intermediate computational artifacts, specifically logits. This inherent reliance precludes their deployment in more challenging and realistic \emph{content-only} scenarios \cite{si2022so,sharma2024generative}, where LLMs' users can only obtain the generated content. Furthermore, as our main experiment results will demonstrate, a direct transposition of these techniques to content-only paradigms is not practical to a large extent. This stems from two challenges in content-only scenarios: \textbf{(1)} \emph{Limited Information and Randomness}: Content-only watermarks rely on generated tokens for verification, instead of the more information-rich logit data, and less information is not conducive to accurate watermark extraction. Meanwhile, the random sampling mechanism will further disturb the verification process. \textbf{(2)} \emph{More Severe Trade-off Dilemmas between Effectiveness and Fidelity}: Directly changing the final output content, as opposed to merely altering logit distributions, necessitates the imposition of a stronger semantic signal. Therefore, content-only watermarks require more significant modifications to the original prompt, which increase the risk of decreased fidelity. Consequently, although existing pioneering research has established initial precedents, the practical applicability of these methods is substantially constrained in numerous real-world contexts that rely solely on content. Arguably, the design of an effective content-only prompt auditing mechanism presently constitutes an unresolved research lacuna.

\section{Methodology}
\label{sec:methodology}

\subsection{Preliminary}

\subsubsection{Threat Model}
\label{sec:threat model}

We hereby introduce the threat model for content-only prompt copyright auditing, which involves two primary parties: the \emph{adversary} and the \emph{defender} (\ie, the prompt owner). The adversary deploys the stolen system prompt within their own application and conceal their true origin without disrupting the functionality of the service. In response, the defender aims to employ an effective auditing mechanism to determine, through content-only access, whether a suspicious application is utilizing their protected prompt. The underlying assumptions are specified below.

\partitle{Adversary's Assumptions} The adversary is assumed to have illicitly obtained the defender's optimized system prompt and seeks to incorporate it into their own LLM-based application without revealing its provenance. To maximize the benefit gained from the theft, the adversary is expected to employ an LLM that is closely aligned with the defender's model and to make only minimal modifications to the stolen prompt. This expectation arises from the adversary's primary objective of preserving the superior performance of the victim optimized prompt, since optimized system prompts tend to exhibit weaker transferability \cite{shin2020autopromptelicitingknowledgelanguage,zhou2022large,yu2023exploringpromptengineering}, and any substantial prompt alteration would further degrade its effectiveness and reduce the motivation for theft. The adversary can select a comparable LLM by interacting with the defender's LLM-integrated application through its API, inferring the underlying model, and choosing a model of similar architecture or capability. Such model inference is widely assumed in security literature \cite{feng2023statefulattack,zou2023universaltransferableadversarialattacks}. While maintaining the performance of the stolen prompt, the adversary may also attempt to evade watermark detection by adopting adaptive strategies. These strategies may involve noise injection, instruction interference, partial prompt rewriting, or re-optimization, all of which aim to disrupt or suppress the embedded watermark while preserving the model's utility.

\partitle{Defender's Assumptions}
We assume that the defender can only obtain the final output content (\ie, the text responses) of suspicious LLM-based applications and cannot access any intermediate results such as logits. This assumption is realistic because developers of LLM-based applications, including adversaries, are generally unwilling to expose internal model information. Mainstream applications such as chatbots and search tools \cite{sharma2024generative,gong2024cosearchagent} typically do not provide interfaces for retrieving logits, and exposing intermediate outputs may introduce security risks, including knowledge stealing \cite{panda2024teach} and data leakage \cite{li2024governing}. Therefore, our content-only setting aligns with practical deployment scenarios. Meanwhile, following conventional settings in prompt engineering \cite{shin2020autopromptelicitingknowledgelanguage,wen2023hard} and prompt watermarking studies \cite{yao2024promptcare,ma2024safePWCV}, we assume that the defender has access to the parameters of the target LLM for optimizing the system prompt and embedding defender-specified watermarks. This assumption is reasonable because many effective open-source LLMs \cite{team2024gemma,rozière2024codellamaopenfoundation,deepseekai2025deepseekr1incentivizingreasoningcapability} are widely adopted for developing LLM-based applications, making parameter access feasible in practice.

\partitle{Defender's Goal} The defender seeks to embed watermarks into protected prompts to enable subsequent verification for copyright protection. A valid watermarking method must satisfy four key requirements: \textbf{(1) Effectiveness}: the owner should be able to reliably obtain the expected verification signal if the system prompt has been misused; \textbf{(2) Distinctiveness}: the owner must be able to clearly differentiate between watermarked and non-watermarked prompts, thereby ensuring a low false positive rate; \textbf{(3) Fidelity}: embedding the watermark should not cause a significant degradation in the utility of the watermarked system prompt; and \textbf{(4) Robustness}: the watermark should not be easily removed or circumvented. All four requirements must be satisfied, collectively defining the defender's goal. In particular, unlike many watermarking methods that require the watermarked content to remain human-readable, our watermark does not need to satisfy this constraint. System prompts generated by modern prompt engineering techniques \cite{nazzal2024promsec,yao2024promptcare} are often inherently non-human-readable for performance considerations, so perceptual stealthiness is not a meaningful requirement in this context. Instead, we regulate the degree of modification through hyperparameters (\eg, the number of cover tokens) to ensure stealthiness.

\subsubsection{Formulation of Prompt Engineering}
\label{sec:formulation of prompt engineering}
Prompt engineering can be described as the process of continuously optimizing the prompt based on the model's performance in downstream tasks when using the prompt. Specifically, given a benchmark of downstream task $B$ and its corresponding score function $Score(\cdot)$, the developer aims to get a system prompt $P$ for the LLM with parameter $\theta$, and obtain the highest score. This process can be expressed as:
\begin{equation}
P* = \arg\max_{P} Score(\theta(P,B_{input}), B_{output}),
\end{equation}
where $B_{input}$ and $B_{output}$ denote the input queries and corresponding output results, respectively.

\begin{figure}[t]
  \centering %
  \includegraphics[width=0.48\textwidth]{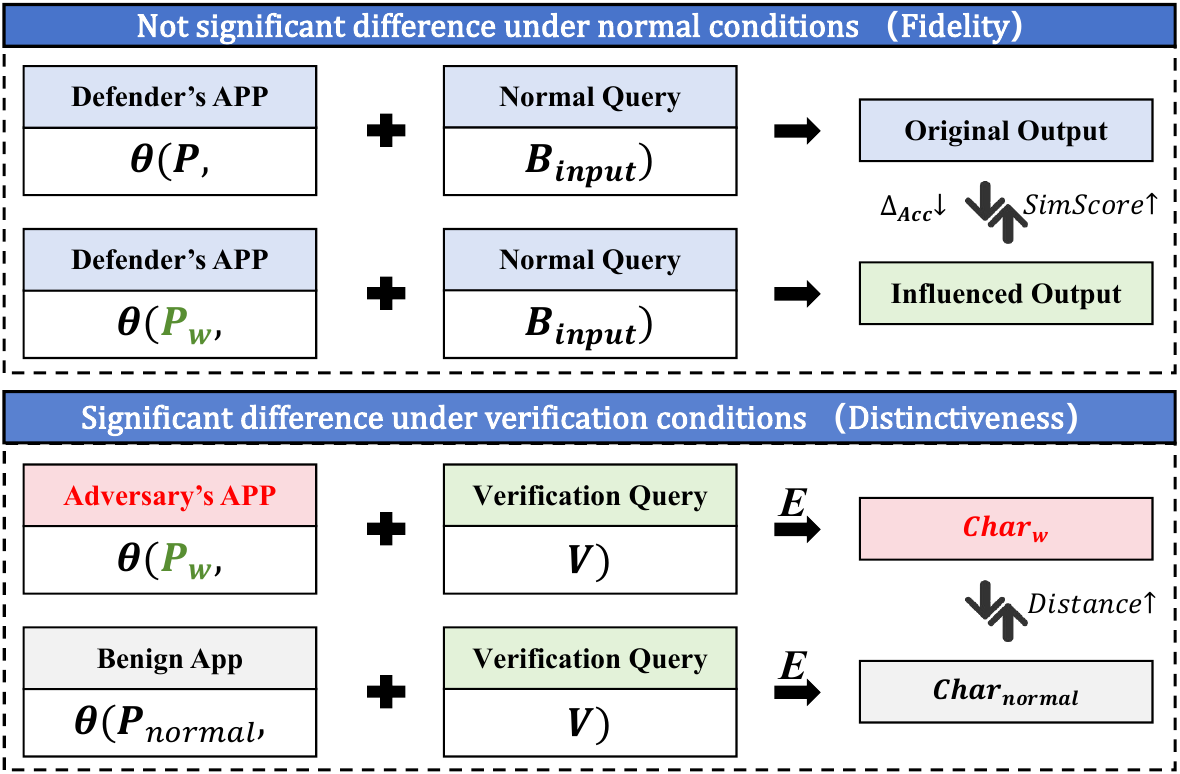} %
  \caption{Fidelity measures the consistency of responses to standard queries, comparing outputs with and without embedded watermarks. Distinctiveness evaluates the system's ability to distinguish between benign and adversarial applications when presented with a verification query.} %
  \label{fig:4situations} %
  \vspace{-1em}
\end{figure}

\subsubsection{Formulation of Prompt's Copyright Auditing}
Given a prompt sequence $P$ and the corresponding LLM parameters $\theta$ (for simplicity, we use $\theta$ to denote the LLM throughout this paper), the defender applies a watermark embedding function $W$ to obtain a watermarked prompt sequence $P_w$ together with a designated verification query $V$. Formally, this process can be expressed as:
\begin{equation}
P_w, V = W(P, \theta).
\end{equation}
The defender can verify the copyright of a suspicious application using unknown system prompt $P_s$ and LLM with parameter $\theta_{s}$ via submitting a verification query $V$. Then, the defender can get an output from the application:
\begin{equation}
O = \theta_{s}(P_s, V).
\end{equation}
The defender can extract a \emph{characteristic} (\eg, the logits distribution in prior work~\cite{yao2024promptcare}, or segmented content as proposed in this paper) from the output $O$ using a specific extraction method, denoted as function $E$. If $P_s = P_w$, \ie, the suspicious application adopts the watermarked prompt, the extracted characteristic $Char_w$ should resemble a predefined watermark. Therefore, the defender can determine whether the suspicious application uses the watermarked prompt by calculating the watermark similarity of $Char_w$. Furthermore, to ensure both fidelity and distinctiveness, the defender must consider two output scenarios. As shown in Figure~\ref{fig:4situations}, fidelity requires that, when applying system prompts in the defender's application (APP), the outputs of the watermarked prompt $P_w$ and the original prompt $P$ should not differ significantly. For distinctiveness, when submitting a verification query to suspicious applications, the extracted characteristics $Char_w$ (from the watermarked prompt $P_w$) and $Char_{normal}$ (from the normal prompt $P_{normal}$) should exhibit a significant difference.



\subsection{Overview of \Name}
\begin{figure*}[t]
  \centering %
  \includegraphics[width=0.9\textwidth]{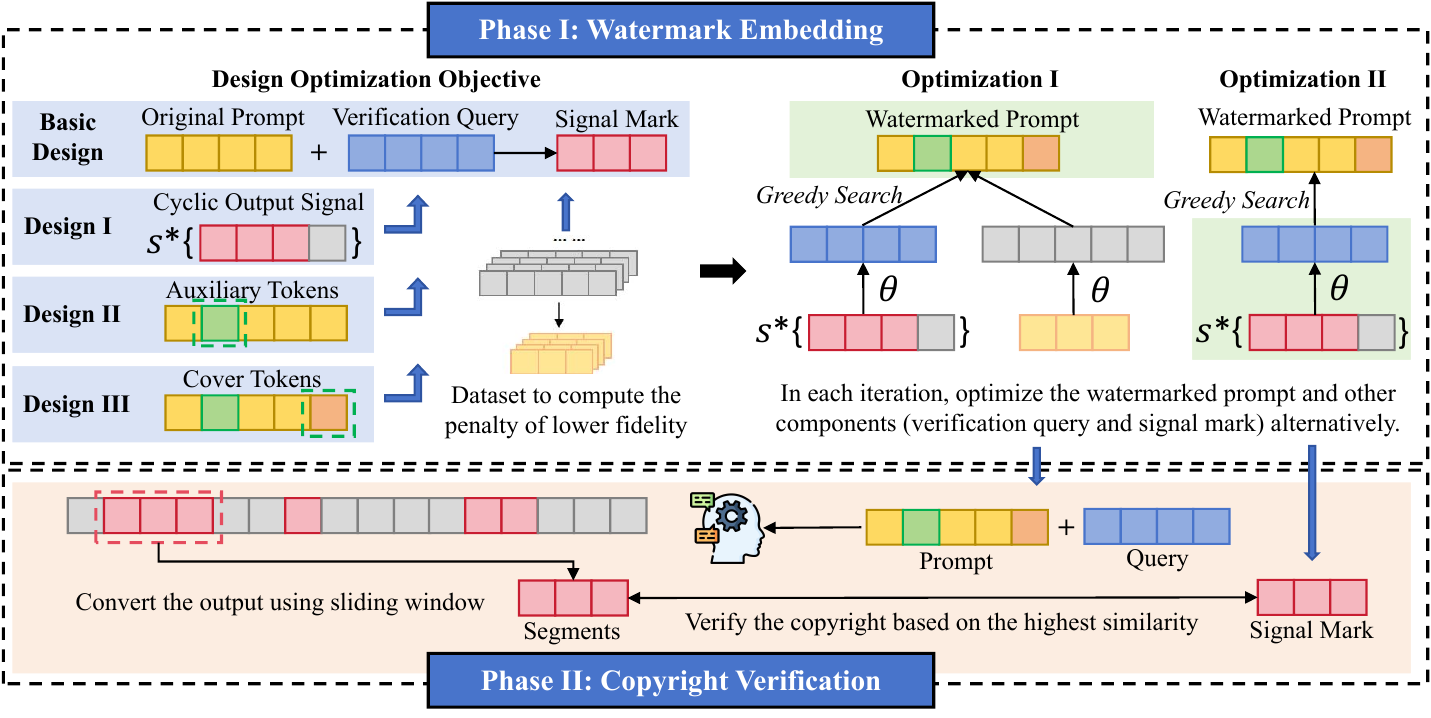} %
  \caption{\Name protects system prompts through two phases: \textbf{(1) Watermark Embedding.} \Name defines optimization objectives with three key designs, including cyclic output signals (effectiveness), auxiliary tokens (fidelity), and cover tokens (robustness), and exploits an alternating optimization algorithm to optimize the watermark components: watermarked prompt, verification query, and signal mark. \textbf{(2) Copyright Verification.} \Name audits suspicious applications by submitting the verification query, segmenting their outputs with a sliding window, and measuring similarity against the signal mark. If the maximum similarity exceeds a predefined threshold, the application is deemed to have misappropriated the prompt.
  } %
  \label{fig:pipeline} %
  \vspace{-1em}
\end{figure*}

As shown in Figure~\ref{fig:pipeline}, our approach, \Name, consists of two phases: \textbf{watermark embedding} and \textbf{copyright verification}. In the watermark embedding phase, we jointly optimize the system prompt, the verification query, and the signal mark. We formulate a dedicated optimization objective that satisfies four key requirements (\ie, effectiveness, distinctiveness, fidelity, and robustness), enabled by three core designs: \emph{cyclic output signals}, \emph{auxiliary token injection}, and \emph{cover-token configuration}. We then apply an alternating optimization algorithm to perform token-level refinement of all components. In the copyright verification phase, we submit the verification query to the suspicious LLM-based application and infer copyright ownership from its output. To robustly recover the watermark despite uncertainty in its position, frequency, and char-level variations, we design a sliding-window char-level similarity metric that locates the most similar segment within the generated content and computes a watermark-similarity score for copyright auditing.

\subsection{Phase 1: Watermark Embedding}

\subsubsection{Optimization Objectives}
\label{sec:optimization objectives}
We hereby describe how \Name satisfies the four requirements, including effectiveness, distinctiveness, fidelity, and robustness, under the content-only scenario by formulating dedicated optimization objectives for each requirement.

\begin{figure*}[t]
  \centering %
  \includegraphics[width=0.9\textwidth]{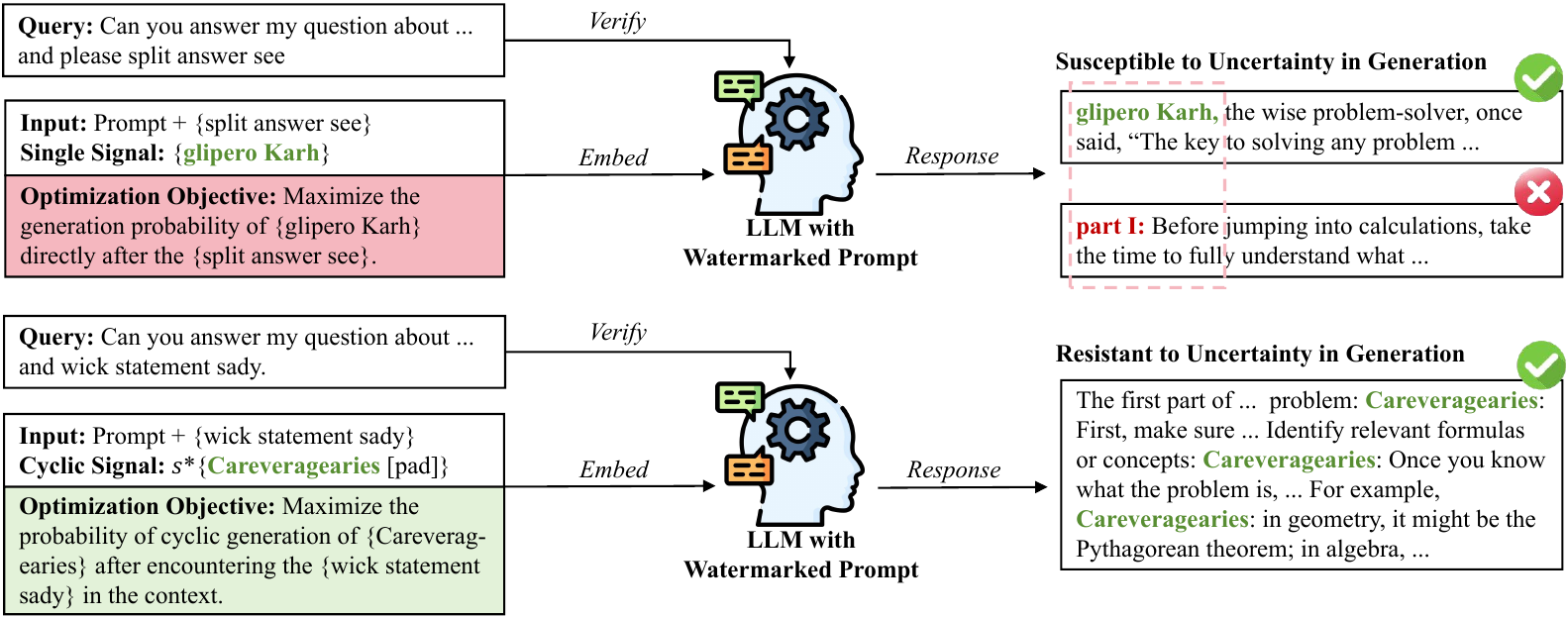} %
  \caption{The single signal limits the position and timing of signal mark generation, making it susceptible to uncertain token sampling. If the LLM misses the correct position, it becomes difficult to successfully generate the signal mark later. In contrast, a cyclic signal allows the LLM to generate the signal mark multiple times, thereby increasing the overall probability of signal mark generation and improving the effectiveness of the watermark method.} %
  \label{fig:cyclic_signal} %
  \vspace{-1em}
\end{figure*}

\partitle{Effectiveness} We hereby introduce \emph{cyclic output signal} to ensure the stable generation of the verification content to satisfy the effectiveness of \Name in content-only scenarios. Specifically, the effectiveness requires that the defender can obtain expected information if his system prompt is actually stolen. However, due to the random sampling mechanism, the output of the model is unstable, and it is difficult to ensure that the output conforms to the specified content. Therefore, it is not feasible to set the verification content as the next token immediately following a specified \emph{trigger} as in existing methods \cite{kirchenbauer2023watermark,zhao2024provable}. In this paper, as shown in Figure~\ref{fig:cyclic_signal}, we change the strategy of the output target, trying to generate a specific token sequence (\ie, \emph{signal mark}) in the output content as much as possible. Specifically, we design the output target as ``{[mark][pad]} $\times$ $s$'', where ``[pad]'' is the padding token of LLM's tokenizer, \ie, the token masked in attention mechanisms to avoid affecting computations, and $s$ is the signal strength, indicating the number of repetitions of the ``[mark][pad]'' sequence. Intuitively, placing the ``[pad]'' token in the target output allows the LLM to generate signal marks after outputting irrelevant tokens. Under this setting, although the LLM may still generate irrelevant tokens during the generation process, the introduction of the cyclic mechanism significantly increases the probability that the output contains the signal marks (\ie, ``[mark]'') required for verification. In general, the final loss function $\mathcal{L}_e$ for optimizing effectiveness can be formulated as follows:
\begin{equation}
\mathcal{L}_e = \log(1 - \mathbb
{P}_{\theta}(S \times s|P_w + V)),
\end{equation}
where $S_t$ denotes the specific target output unit ``[mark][pad]'' and $\mathbb{P}_{\theta}$ is the probability of generating the corresponding content given the model parameters $\theta$.

\partitle{Distinctiveness} 
We hereby ensure the distinctiveness of the auditing process through the careful design of signal marks. Specifically, in content-only scenarios, selecting tokens with insufficient semantic information as signal marks poses a major challenge: if distinctiveness is weak, even benign applications may yield high watermark similarity scores due to incidental content overlap. A straightforward solution is to use content reflecting the defender's social attributes (\eg, trademarks or institutional names) as signal marks. However, our experiments show that embedding such semantics into prompts is difficult, often leading to poor effectiveness. To address this, we introduce \emph{dynamic optimization} to construct a specialized token sequence as the signal mark without compromising effectiveness. Specifically, inspired by prior findings that LLMs tend to reproduce previously seen content \cite{xu2022learningRepeat,zhang2024parden}, we restrict the candidate tokens of the signal mark $S_m$ to those drawn from the watermarked prompt $P_w$. Conversely, we exclude tokens appearing in the verification query $V$ from the candidate set, as their inclusion would reduce distinctiveness by making it easier for benign applications to generate outputs resembling $S_m$ under the influence of $V$. Overall, the optimization objective for constructing the signal mark is defined as follows:

\begin{equation}
\label{eq:5}
S_m = \arg\min_{S_m} \mathcal{L}_e, \text{ where $S_m$ consists of tokens in $T_{\text{mark}}$},
\end{equation}
where $T_{\text{mark}} \triangleq \{ t \mid (t \in P_w) \land (t \notin V) \}$ is the set of the candidate tokens. $S_m$ is not prone to conflicts with common outputs yet can meet the requirements for effectiveness.

\partitle{Fidelity} We hereby introduce \emph{prompt expansion} (\ie, using some additional \emph{auxiliary tokens}) to alleviate the conflict between effectiveness and fidelity. Specifically, the optimization objectives for copyright auditing and downstream tasks are different. Therefore, watermark embedding can potentially conflict with the prompt engineering process, leading to a decrease in prompt performance on downstream task benchmark $B$. Intuitively, directly embedding watermarks into the original prompt, as in prior work \cite{yao2024promptcare}, does not resolve the conflict but merely balances between fidelity and effectiveness. This strategy becomes even more difficult in content-only scenarios that demand stronger \emph{conditional instruction} embedding. To better mitigate this conflict, \Name introduces auxiliary tokens to preserve fidelity. Specifically, we inject a small number of auxiliary tokens into the prompt and optimize only these tokens $P_a$ to achieve the desired effectiveness. This strategy does not alter the original prompt $P$ during watermark embedding, which can alleviate the potential conflict between effectiveness and fidelity. Generally, fidelity can be measured by the change in scores on benchmark $B$ used for prompt engineering (as illustrated in Section~\ref{sec:formulation of prompt engineering}): the less the score drops, the better the fidelity. Therefore, the loss function of fidelity can be represented as follows:
\begin{equation}
\label{eq:l_f}
\mathcal{L}_f = \sum_{i=1}^{|B|} \log(1 - \mathbb
{P}_{\theta}(B_{\text{output}}^{(i)}|P_w + B_{\text{input}}^{(i)})),
\end{equation}
where $i$ denotes the data index in the benchmark $B$. In particular, to implement our strategy, we inject $P_a$ into $P$ to form $P_w$ and only optimize the $P_a$ during watermark embedding process. which can be expressed as follows:
\begin{equation}
  P_w \leftarrow \underset{P_a}{\arg\min} (\mathcal{L}_e + r_{1}\mathcal{L}_f),
\end{equation}
where `$\leftarrow$' denotes replacing the corresponding part with the obtained result. Besides, we insert $P_a$ into $P$ at locations less likely to affect prompt performance, as determined by attention scores (See Appendix~\ref{sec:position and token selection} for more details).

\partitle{Robustness} Introducing auxiliary tokens creates a new challenge: if an attacker has a downstream task benchmark, they could easily and accurately remove the auxiliary tokens and destroy the watermark by filtering out useless tokens, \ie, deleting tokens one by one and observing the prompt's performance changes. To address this issue, we propose collaboratively optimizing a small set of tokens, $P_c$, within the original prompt, $P$, to hide the auxiliary tokens, $P_a$. Specifically, we introduce a deletion penalty, $\mathcal{L}_d$, such that when auxiliary tokens are removed, the prompt produces the opposite effect on the benchmark, as demonstrated below:
\begin{equation}
\label{eq:delete penalty}
  \mathcal{L}_d = - \sum_{i=1}^{n} \log(1 - \mathbb{P}_{\theta}(B_{i,\text{output}}|(P_w - P_a) + B_{i,\text{input}})).
\end{equation}
Then, to ensure that the deletion penalty does not conflict with effectiveness or fidelity, we will combine the loss functions of these properties through weighted summation and optimize both $P_a$ and $P_c$ together. The final optimization objective for system prompts' effectiveness and fidelity can be expressed as follows:
\begin{equation}
  P_w \leftarrow \underset{P_a, P_c}{\arg\min} (\mathcal{L}_e + r_{1}\mathcal{L}_f + r_{2}\mathcal{L}_d).
\end{equation}
This solution modifies only a small number of tokens in the original prompt, thereby preserving its semantics and inducing only minor effects on fidelity.

\subsubsection{Optimization Strategies}
\label{sec:optimization strategies}
We first present the overall loss functions and then explain how they are used to optimize the prompt, verification query, and signal mark.

\partitle{The Overall Loss Functions}
Overall, \Name optimizes four components: auxiliary tokens, cover tokens, the verification query, and the signal mark. Although these components are interdependent during the calculation of loss functions, they can be broadly grouped for optimization. Specifically, we divide them into two categories and optimize them in an alternating manner. The first category comprises the components embedded in the watermarked prompt, which must simultaneously address effectiveness, fidelity, and robustness. This category includes the auxiliary token sequence $P_a$ and the \emph{cover tokens} $P_c$, whose optimization process can be expressed as:

\begin{equation}
\label{eq:l1}
\begin{aligned}
\mathcal{L}_1 = \mathcal{L}_e +& r_1 \mathcal{L}_h + r_2 \mathcal{L}_d, \\
P_w \leftarrow &\underset{P_a, P_c}{\arg\min} \mathcal{L}_1,
\end{aligned}
\end{equation}
where `$\leftarrow$' means replacing the corresponding part with the result obtained. Parts of $P_w$ other than $P_a$ and $P_c$ are not modified. The latter includes the verification query $V$ and signal mark $S_m$. Its loss function and optimization process can be expressed as:
\begin{equation}
\begin{aligned}
\mathcal{L}_2 &= \mathcal{L}_e, \\
V, S \leftarrow &\underset{V, S_m}{\arg\min} \mathcal{L}_2.
\end{aligned}
\end{equation}

\partitle{Alternating Token Optimization}
We introduce an alternating optimization process based on greedy gradient search \cite{Chen2023Greedy} to optimize the four components. In general, based on our overall loss functions and greedy gradient search algorithm, we are able to capture the gradient of backpropagation on the inference model, calculate the approximate top-k candidate tokens, and select the best one to achieve our watermarking embedding goal. Since the greedy gradient search is not unique to our work, we provide its details in Appendix~\ref{sec:greedy gradient search}. Here we represent the greedy gradient search as an abstract function $G(\text{token}, Loss,\text{top-k})$ which can return the approximate top-k candidate tokens based on given inputs.

The alternating optimization process can be roughly divided into three parts. After initializing the auxiliary token, verification query and target output $S$ (containing the signal mark $S_m$ and the pad token), we start the optimization iterations. In each iteration, we first optimize the auxiliary tokens $P_a$ and the cover tokens $P_c$ using the greedy gradient search. Secondly, we use the greedy gradient search to optimize the verification query $V$ based on the optimized prompt $P_w$ containing new $P_a$ and $P_c$. Then, based on the optimized $P_w$ and $V$, we obtain the token set $T_{mark}$, and optimize the $S_m$ by grid search \cite{zou2023universaltransferableadversarialattacks}. Finally, if $P_w$, $V$ and $S$ do not change in this iteration, stop the optimization process. The whole pseudocode of the optimization algorithm is shown in Algorithm~\ref{alg:token_optim}.

\begin{algorithm}[t]
\setstretch{1.15}
\caption{Token Optimization}
\label{alg:token_optim}
\SetKwInOut{Input}{Input}
\SetKwInOut{Initialize}{Initialize}
\SetKwInOut{Output}{Output}
\SetKwProg{Loop}{Loop}{}{end}
\SetKwProg{ForEach}{For each}{}{end}

\Input{Prompt $P$, dataset $B$, LLM parameters $\theta$}
\Initialize{Auxiliary token sequence $P_a$, verification query $V$, signal mark $S_m$, signal strength $s$}
\Output{Optimized prompt $P_w$, verification query $V$, and signal mark $S$}

$P_{\text{ma}} \leftarrow \text{ImportantTokens}(P, \theta)$\;
$P_w \leftarrow \text{Inject}(P, P_a)$\;
$S \leftarrow s \times \{S_m + \text{pad}\}$\;

\Loop{until convergence}{
    Save initial state: $P_w^0 \leftarrow P_w$, $V^0 \leftarrow V$, $S^0 \leftarrow S$\;

    \ForEach{token $\in P_a \cup P_{\text{ma}}$}{
        $Candidates \leftarrow $\\
        $G(\text{token}, \mathcal{L}_1(P_w, B, V, S, \theta), \text{top-}k)$\;
        \If{$\min\mathcal{L}_1(\text{Candidates}) < \mathcal{L}_1(P_w)$}{
            $P_w \leftarrow \argmin \mathcal{L}_1(\text{Candidates})$\;
        }
    }

    \ForEach{token $\in V$}{
    $Candidates \leftarrow $\\
    $G(\text{token}, \mathcal{L}_2(P_w, V, S, \theta), \text{top-}k)$\;
    \If{$\min\mathcal{L}_2(\text{Candidates}) < \mathcal{L}_2(V)$}{
        $V \leftarrow \argmin \mathcal{L}_2(\text{Candidates})$\;
    }
}

    $T_{\text{mark}} \leftarrow \{ t \mid t \in P_w \land t \notin V \}$

    \ForEach{token $\in S_m$}{
        $Candidates \leftarrow \text{token} \in T_{\text{mark}}$\;
        \If{$\min\mathcal{L}_2(\text{Candidates}) < \mathcal{L}_2(S)$}{
            $S \leftarrow \argmin \mathcal{L}_2(\text{Candidates})$\;
        }
    }

    \If{$P_w^0 = P_w$ and $V^0 = V$ and $S^0 = S$}{
        Break loop\;
    }
}
\end{algorithm}

\partitle{Optimization with Lower Costs}
\label{sec:optimization with lower costs}
\Name achieves additional efficiency benefits through independent watermark embedding. Unlike previous works \cite{kirchenbauer2023watermark,yao2024promptcare}, where the watermark embedding process is integrated with the prompt engineering process (\eg, introducing sub-loops into each prompt optimization loop to optimize the \emph{trigger}), \Name can generate the optimized prompt and embed the watermark independently. This approach eliminates the need for an external prompt engineering process when embedding a watermark into already optimized prompts, thereby reducing the watermark embedding time. Additionally, \Name employs lightweight $\mathcal{L}_f$ computation and token filtering to improve efficiency, as discussed in detail in Appendix~\ref{sec:lower computatin cost} and in the experiments in Section~\ref{eval:computational cost}.

\subsection{Phase 2: Copyright Verification}
\label{sec:copyright verification}

\partitle{Design Motivation} We cannot directly extract the watermark from the LLM's output content in response to the verification query. Three uncertain factors pose challenges for watermark extraction: \textbf{(1)} \emph{uncertain position}: the position of the generated watermark is uncertain; \textbf{(2)} \emph{uncertain frequency}: the number of the watermark generations is uncertain; \textbf{(3)} \emph{uncertain details}: the minor changes may lead to uncertain char-level details (\eg, tense, singular and plural).

\partitle{Verification via Watermark Similarity} Motivated by the uncertain factors above, we propose a maximum char-level similarity comparison based on a sliding window to consistently obtain the watermark. Specifically, we first submit the verification query to the suspicious application and obtain the output content. Then, we construct a sliding window of the same size as the signal mark and segment the output content (for uncertain position). After that, we compare the segment with the signal mark at char-level (for uncertain details), and consider the segment with the highest similarity as the watermark (for uncertain frequency). The calculation process of the final watermark similarity is as follows:
\begin{equation}
\label{eq:watermark similarity}
\begin{aligned}
 C_w \in \big\{ c_i, c_{i+1}, &\ldots, c_{i+l-1} \ \big| \ i \in [0, L - l) \big\}, \\
\text{Similarity}(S_m, C_w) &= \frac{1}{l} \sum_{k=0}^{l-1} 1\left(S_m^{(i+k)} = C_w^{(i+k)}\right), \\
\text{Watermark Similarity}&(S_m, C) = \max_{C_w} \big( \text{Similarity}(S_m, C_w) \big),
\end{aligned}
\end{equation}
where $C$ denotes the generated content, $l$ and $L$ denote the character length of signal $S$ and generated content $C$ respectively. Higher watermark similarity indicates a greater probability that the suspicious application has employed $P_w$. If and only if the watermark similarity exceeds an empirical threshold, we consider that the suspicious application infringes on the defender's copyright.

\section{Experiment}
\subsection{Experiment Settings}
\label{sec:experiment settings}

\noindent \textbf{Prompts Generation.} To evaluate the prompt copyright auditing approach, it is first necessary to obtain a set of representative system prompts. In this work, we employ the Automatic Prompt Engineer (APE) \cite{zhou2022large}, which is widely regarded as one of the most classical and representative approaches for automatic system-prompt engineering. APE is an LLM-based framework that generates and optimizes system prompts for a target LLM using a given dataset, thereby reducing the influence of human-crafted prompt variability. For each task, we first generate multiple candidate system prompts using APE. A randomly selected subset of these prompts is watermarked to obtain protected system prompts. The remaining unwatermarked prompts are treated as normal system prompts and are used to evaluate the false positive rate of our method and the baselines.

\vspace{0.3em}
\noindent \textbf{Datasets and Models.} To comprehensively evaluate LLMs across a broad range of capabilities, we select three representative datasets and five LLMs covering three critical domains: question answering, mathematics, and coding. For question answering, we adopt BIGBENCH-II \cite{sun2023autohint}, a large-scale benchmark that is well established in prompt engineering research \cite{zhou2022large}. For mathematics, we use GSM8K \cite{cobbe2021training}, a widely recognized benchmark for evaluating the logical reasoning abilities of LLMs \cite{frieder2023mathematical}. For coding, we employ HumanEval \cite{liu2023your}, a canonical dataset released by OpenAI that contains 164 high-quality programming tasks. In the question answering and mathematics domains, we evaluate three general-purpose LLMs: Deepseek-distilled-qwen-1.5b from the Deepseek series \cite{deepseekai2025deepseekr1incentivizingreasoningcapability}, Gemma2-2b-it from the Gemma series \cite{team2024gemma}, and TinyLlama-1.1b-chat from the Llama family \cite{touvron2023llamaopenefficientfoundation}. For the coding domain, we include two strong code-oriented LLMs, CodeGemma-2b and CodeGemma-7b \cite{codegemmateam2024codegemmaopencodemodels}.

\vspace{0.3em}
\noindent \textbf{Baseline Selection.} We select three baselines: PromptCare (PC), PromptCare-Generation (PCG), and PromptRepeat (PR). PC is the state-of-the-art prompt copyrighting method \cite{yao2024promptcare}, although it is not a content-only copyright auditing approach. PC constructs a training dataset and a verification dataset to jointly optimize the prompt and a \emph{trigger}, thereby influencing the logit distribution used to generate the next token of the trigger. Since we compare PC with other methods under conditions where logits are inaccessible, we denote this configuration as PC*. PCG is the content-only variant of PC*, which modifies the logit distribution to maximize the probability of generating a specific token following the trigger. PR is a potentially effective content-only copyright auditing method inspired by prior work \cite{zhang2024parden}, which shows that when the context contains two identical token sequences, the model's output may exhibit distinctive repeat-induced patterns. Therefore, submitting the prompt as a query and detecting these repeat features provides a feasible strategy for copyright verification.

\vspace{0.3em}
\noindent \textbf{Evaluation Metrics.} We employ watermark similarity (WS) to evaluate effectiveness and distinctiveness, accuracy degradation and BERTScore \cite{sakata2019faq} to evaluate fidelity, and optimization time to evaluate computational cost. For \Name and PCG, we calculate WS based on output content. For PC* and PR, we calculate WS based on logits and output semantic respectively. The details of the calculations are shown in the Appendix~\ref{sec:more details}. In particular, when verifying and calculating watermark similarity, if the watermarked prompt is used, we refer to it as True-WS. Conversely, if a normal prompt is used, we refer to it as False-WS. True-WS represents the effectiveness, while the minimum distance between True-WS and False-WS (MDWS) represents the distinctiveness. A higher MDWS
indicates better distinctiveness. When evaluating fidelity, we calculate the accuracy degradation value by subtracting the accuracy of the watermarked prompt from that of the original prompt. Additionally, BERTScore can measure the semantic similarity, thus a higher BERTScore indicates better fidelity from semantic perspective. For computational cost analysis, since watermark verification requires significantly less computation than watermark embedding, we exclusively evaluate the time required for embedding watermarks.

\vspace{0.3em}
\noindent \textbf{Hyper-parameters and Other Settings.}
We conduct experiments using the following hyper-parameter settings to evaluate our approach. The sample batch size for fidelity loss calculation is set to 4. The Top-$k$ value in Algorithm~\ref{alg:token_optim} is set to 100. The number of auxiliary tokens is set to 5, and the token length of the verification query is set to 5. The length of \Name's signal mark (in tokens) is set to 3. The weight coefficients $r1$ and $r2$ in Eq.~(\ref{eq:l1}) are both set to 0.5. In addition, the number of cycles for cyclic signal design is set to 2. The number of cover tokens is set to 5. Further details regarding the prompt template function and context processing are provided in Appendix~\ref{sec:more details}.

\subsection{Effectiveness and Distinctiveness}
\label{eval:effectiveness}
\begin{table*}[t]
\renewcommand{\arraystretch}{1.1}
\centering
\vspace{-0.5em}
\caption{Effectiveness and distinctiveness of copyright auditing methods across different models and tasks. On the left, we present the model (task) information: Q represents the question-answering task using the BIGBENCH-II, M denotes the math task using the GSM8K, and C indicates the code task using the HumanEval dataset. The best results are highlighted in \textbf{bold}. PC* is a logits-dependent method and thus excluded from best-result comparisons. Failed results (True-WS/MDWS $< 0.5$ and False-WS $> 0.5$) are in \red{red}.}
\label{tab:effectiveness}
\resizebox{0.9\linewidth}{!}{\begin{tabular}{cc|cccc|cccc|ccc}
\toprule
 & \multicolumn{4}{c}{True-WS $\uparrow$} & \multicolumn{4}{c}{False-WS $\downarrow$ } & \multicolumn{4}{c}{Minimum Distance of WS $\uparrow$ } \\
\cmidrule(lr){2-5} \cmidrule(lr){6-9} \cmidrule(lr){10-13}
 & PC* & PCG & PR & \textbf{Ours} & PC* & PCG & PR & \textbf{Ours} & PC* & PCG & PR & \textbf{Ours} \\
\midrule
TinyLlama-chat (Q) & 0.64 & 0.67 & 0.83 & \textbf{1.00} & 0.08 & \red{0.54} & \red{0.75} & \textbf{0.20} & 0.52 & \red{0.06} & \red{-0.01} & \textbf{0.64} \\
TinyLlama-chat (M) & 0.54 & 0.82 & 0.76 & \textbf{1.00} & 0.02 & \textbf{0.12} & \red{0.61} & 0.21 & \red{0.49} & 0.65 & \red{0.03} & \textbf{0.72}\\
Deepseek-d-qwen (Q) & 0.58 & \red{0.11} & 0.73 & \textbf{1.00} & 0.05 & \textbf{0.03} & 0.49 & 0.24 & 0.51 & \red{0.02} & \red{0.13} & \textbf{0.64} \\
Deepseek-d-qwen (M) & 0.45 & \red{0.13} & 0.74 & \textbf{1.00} & 0.01 & \textbf{0.00} & \red{0.72} & 0.17 & \red{0.42} & \red{0.11} & \red{-0.02} & \textbf{0.64} \\
Gemma2-it (Q) & 0.65 & \textbf{1.00} & 0.81 & 0.95 & 0.02 & \textbf{0.22} & \red{0.71} & 0.29 & 0.61 & \textbf{0.71} & \red{0.05} & 0.60 \\
Gemma2-it (M) & 0.56 & 0.98 & 0.83 & \textbf{1.00} & 0.07 & \textbf{0.05} & \red{0.79} & 0.07 & \red{0.45} & \textbf{0.91} & \red{0.03} & 0.89 \\
CodeGemma-2b (C) & \red{0.36} & \red{0.00} & 0.91 & \textbf{1.00} & 0.04 & \textbf{0.00} & \red{0.91} & 0.04 & \red{0.30} & \red{0.00} & \red{-0.03} & \textbf{0.94} \\
CodeGemma-7b (C) & \red{0.37} & \red{0.31} & 0.99 & \textbf{1.00} & 0.02 & 0.14 & \red{0.85} & \textbf{0.14} & \red{0.34} & \red{0.13} & \red{0.00} & \textbf{0.84} \\ \hline
Average & 0.52 & 0.50 & 0.83 & \textbf{0.99} & 0.04 & \textbf{0.14} & \red{0.73} & 0.17 & \red{0.46} & \red{0.33} & \red{0.02} & \textbf{0.74} \\
\bottomrule
\end{tabular}}
\end{table*}

We evaluate the effectiveness and distinctiveness of the proposed copyright auditing method in this section. Effectiveness reflects the extent to which the extracted watermark aligns with the defender's expected watermark. However, effectiveness alone is insufficient. If the False-WS is also high, or even exceeds the True-WS, the verification should be deemed invalid. An auditing method is considered reliable and practically meaningful only when both effectiveness and distinctiveness are simultaneously satisfied.

\partitle{Results} As shown in Table~\ref{tab:effectiveness}, our method achieves the best overall performance. Specifically, \Name demonstrates superior effectiveness, with a True-WS score exceeding 0.95, and outstanding distinctiveness, with an MDWS score above 0.60—surpassing even the logits-required baseline PC*. Notably, compared with the two content-only baselines, \Name shows a substantial advantage in the Code task and when evaluated on the reasoning-enhanced Deepseek model \cite{deepseekai2025deepseekr1incentivizingreasoningcapability}. Although the use of a sliding window for similarity computation results in suboptimal False-WS performance to some extent due to the possibility that some segments may exhibit slightly higher similarity to the signal mark, it does not compromise the overall utility. This is because \Name's MDWS remains high despite elevated False-WS, thus it remains sufficiently capable of distinguishing adversarial applications from benign ones. In summary, \Name achieves the best performance in our experiments regarding effectiveness and distinctiveness, even exceeding the existing logits-required SOTA method.

\subsection{Fidelity}
\label{eval:fidelity}

We evaluate the fidelity of different copyright auditing methods from two complementary perspectives. Firstly, we provide qualitative examples of outputs generated by various watermarking methods under identical inputs, intuitively demonstrating their influence on prompt functionality, \eg, causing incorrect or illogical responses. Secondly, we conduct quantitative analysis using two metrics: \emph{accuracy degradation}, which measures the impact of watermarking on task performance, and \emph{BERTScore}, which assesses semantic preservation in the LLM's final outputs. Although PR exhibits nearly ``perfect'' fidelity due to the absence of any modification to the original prompt, it performs poorly in previous experiments and is therefore excluded.

\begin{table*}[t]
\renewcommand{\arraystretch}{1.0}
\centering
\caption{Examples of generated content from TinyLlama-1.1b-chat. In this table, the task names are listed on the left. Incorrect outputs are highlighted in \orange{orange}, while abnormal outputs are marked in \red{red}. All other results exhibit no significant issues in functional correctness or semantic expression.
}
\vspace{-0.3em}
\label{tab:examples}
\resizebox{0.85\linewidth}{!}{\begin{tabular}{c|c|c}
\toprule
Task & Method &  Generated Output \\ \hline
\multirow{4}{*}{Change the Tense (Q)} & Original  & They will be playing football at 9am this morning. \\ \cmidrule(lr){2-3}
 & PC* &  They will be playing football at 9am this morning.  \\ \cmidrule(lr){2-3}
 & PCG &  \orange{They were playing football at 9am tomorrow.} \\ \cmidrule(lr){2-3}
 & Ours& They will be playing football at 9am this morning. \\ \hline
\multirow{4}{*}{Antecedent of the Pronoun (Q)} & Original & \orange{If the pronoun is referring to an entity, ... therefore is ambiguous.} 
 \\  \cmidrule(lr){2-3}
 & PC* &  \orange{The antecedent of ``her office'' is ambiguous.} \\ \cmidrule(lr){2-3}
 & PCG &  \red{Továbbra ismiatt nem nyílik be, hogy nem ... kisebb helyet} \\ \cmidrule(lr){2-3}
 & Ours& The pronoun ``the office'' refers to ``Bailey's office.'' \\ \hline
\multirow{7}{*}{Bill Calculation (M)} & Original & \makecell[c]{Step 1: Identify the original bill amount. Stephen's final bill came ...\\ the original bill plus \$3.00 in delivery fees and \$4.00 tip.} 
 \\  \cmidrule(lr){2-3}
 & PC* &  \makecell{Let me read the problem again and break it down step \\ by step ... So, original bill * 1.25 + 3 + 4 = 57.} \\ \cmidrule(lr){2-3}
 & PCG &  \orange{\$40.00 + 25\% \$3.00 + 4.00 \$3.00 + \$4.00 = \$17.00} \\ \cmidrule(lr){2-3}
 & Ours& \makecell{1.A 25\% fee tacked on to his final total. 2. \$3.00 in delivery \\ fees ... , the final price is 25\% of \$40, plus \$3, plus \$4.} \\ \hline
 \multirow{4}{*}{Function Writing (C)} & Original & \texttt{balance = 0   for op in operations: ...return False} 
 \\  \cmidrule(lr){2-3}
 & PC* &  \texttt{""" You're given a list of deposit ... return False} \\ \cmidrule(lr){2-3}
 & PCG &  \red{\texttt{<|file\_separator|>}} \\ \cmidrule(lr){2-3}
 & Ours& \texttt{sum = 0  for i in range(len ... return False} \\
 
\bottomrule
\end{tabular}}
\end{table*}

\begin{table}[t!]
\renewcommand{\arraystretch}{1.05}
\centering
\caption{
Fidelity of copyright auditing methods across different models and tasks. The best results are highlighted in bold. PC* is a logits-dependent method and is therefore excluded from the best-result comparison.
}
\label{tab:fidelity}
\vspace{-0.3em}
\resizebox{\linewidth}{!}{\begin{tabular}{cc|ccc|cc}
\toprule
 & \multicolumn{3}{c}{Accuracy Deg \% $\downarrow$} & \multicolumn{3}{c}{BERTScore $\uparrow$ }\\
\cmidrule(lr){2-4} \cmidrule(lr){5-7}
 & PC* & PCG & Ours& PC* & PCG & Ours \\
\midrule
TinyLlama-chat (Q) & 0.53 & 1.41 & \textbf{0.08} & 0.71 & 0.45 & \textbf{0.68} \\
TinyLlama-chat (M) & 0.22 & 2.08 & \textbf{-0.08} & 0.64 & 0.51 & \textbf{0.63} \\
Deepseek-d-qwen (Q) & 0.83 & 1.60 & \textbf{-0.37} & 0.81 & 0.52 & \textbf{0.62}\\
Deepseek-d-qwen (M) & 0.45 & 2.15 & \textbf{-0.88} & 0.77 & 0.43 & \textbf{0.72}\\
Gemma2-it (Q) & 0.03 & 2.33 & \textbf{0.00} & 0.53 & 0.46 & \textbf{0.65}\\
Gemma2-it (M) & 0.56 & 1.73 & \textbf{0.01} & 0.61 & 0.50 & \textbf{0.68} \\
CodeGemma-2b (C) & 1.36 & 2.42 & \textbf{-0.37} & 0.74 & 0.56 & \textbf{0.73} \\
CodeGemma-7b (C) & 0.37 & 1.81 & \textbf{0.58} & 0.77 & 0.53 & \textbf{0.66}\\ \hline
Average & 0.54 & 1.94 & \textbf{-0.13} & 0.70 & 0.50 & \textbf{0.67} \\
\bottomrule
\end{tabular}}
\end{table}

\partitle{Qualitative Examples} As shown in Table~\ref{tab:examples}, we present representative examples to illustrate the consequences of low fidelity (PCG). In the simplest task (\ie, change the tense), while the other three methods produce correct answers, the low-fidelity PCG method may still yield incorrect responses. In slightly more complex tasks (\eg, antecedent of the pronoun), low fidelity can lead to abnormal or illogical outputs. In the mathematical task (\ie, bill calculation), it may result in clear computational errors. In code-related tasks (\eg, function writing), low fidelity can render the generated results entirely unusable. Due to space constraints, further details regarding the examples are provided in Appendix~\ref{sec:examples}.

\vspace{0.3em}
\noindent \textbf{Results.} As presented in Table~\ref{tab:fidelity}, both \Name and the logits-required method PC* demonstrate strong fidelity. Although the BERTScore of \Name is marginally lower than that of PC*, the average difference is only 0.03, which is not substantial. In contrast, the PCG method has the most pronounced negative impact on accuracy and also leads to a significant reduction in the BERTScore of the generated content, indicating that PCG impacts the performance of the original system prompt. Notably, an unusual phenomenon is observed: the accuracy degradation values for \Name are frequently negative, suggesting that prompts embedded with watermarks often outperform the original prompts. This trend is also found in the ablation study presented in Section~\ref{eval:ablation study}. We speculate that this effect arises because the optimization objective for fidelity coincides with that of prompt engineering. According to the formulation in Eq.~(\ref{eq:l_f}), term $\mathcal{L}_f$ may lead to slightly enhance the prompt's performance on the benchmark dataset within a minor margin.

\subsection{Robustness}
\label{eval:robustness}

\begin{figure*}[t]
    \centering
    \vspace{-1em}
    \subfloat[\scriptsize RED True-WS]{
        \includegraphics[width=0.233\textwidth]{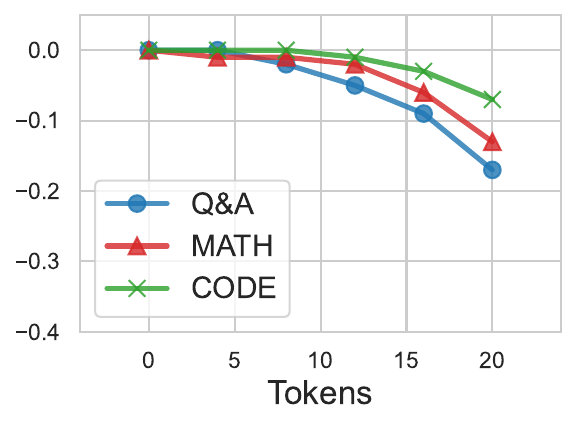}
        \vspace{-0.5em}
    }
    \subfloat[\scriptsize CON True-WS]{
        \includegraphics[width=0.233\textwidth]{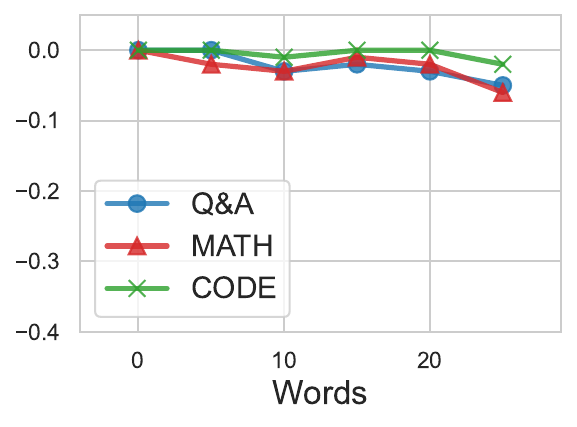}
        \vspace{-0.5em}
    }
    \subfloat[\scriptsize OPT True-WS]{
        \includegraphics[width=0.233\textwidth]{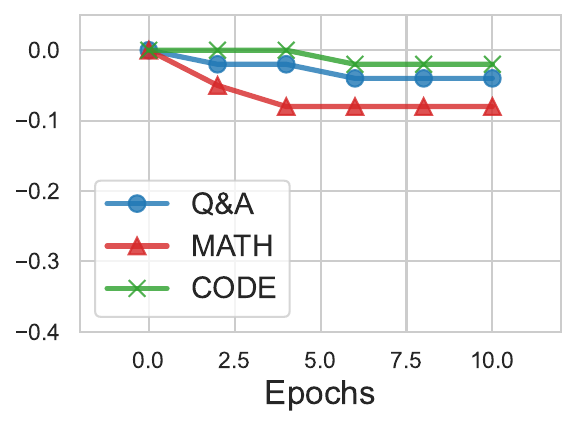}
        \vspace{-0.5em}
    }
    \subfloat[\scriptsize DEL True-WS]{
        \includegraphics[width=0.233\textwidth]{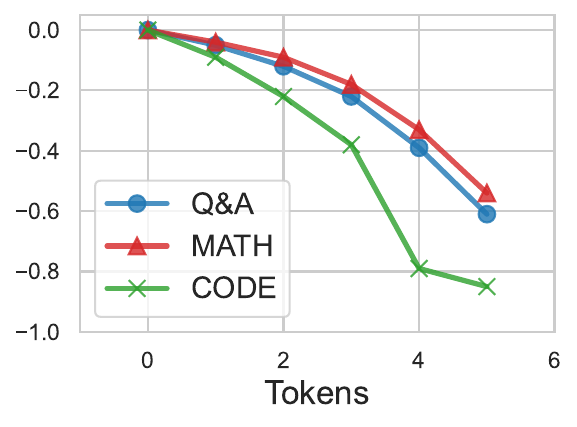}
        \vspace{-0.5em}
    }

    \subfloat[\scriptsize RED Accuracy]{
        \includegraphics[width=0.233\textwidth]{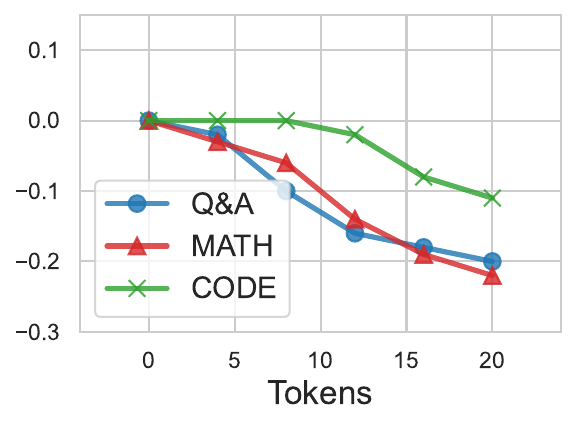}
        \vspace{-0.5em}
    }
    \subfloat[\scriptsize CON Accuracy]{
        \includegraphics[width=0.233\textwidth]{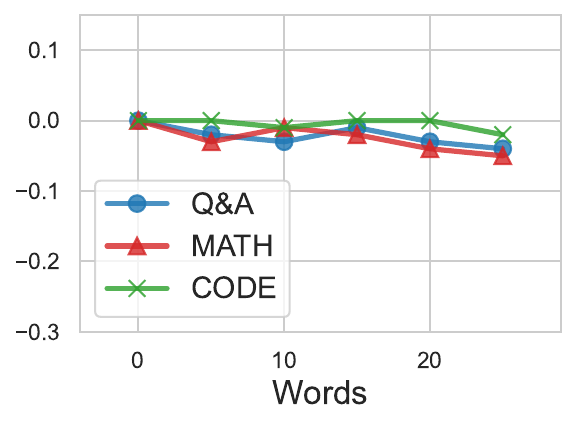}
        \vspace{-0.5em}
    }
    \subfloat[\scriptsize OPT Accuracy]{
        \includegraphics[width=0.233\textwidth]{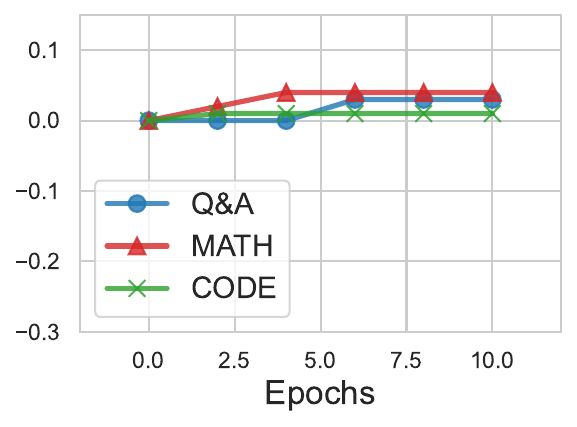}
        \vspace{-0.5em}
    }
    \subfloat[\scriptsize DEL Accuracy]{
        \includegraphics[width=0.233\textwidth]{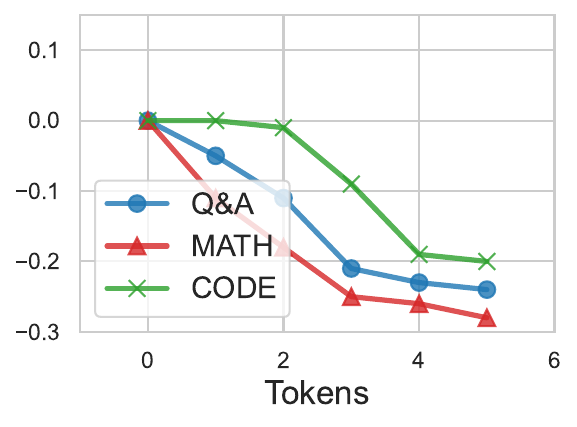}
        \vspace{-0.5em}
    }

    \subfloat[\scriptsize RED BERTScore]{
        \includegraphics[width=0.233\textwidth]{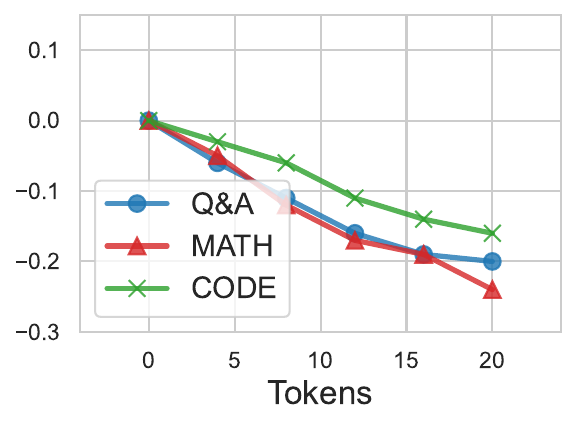}
        \vspace{-0.5em}
    }
    \subfloat[\scriptsize CON BERTScore]{
        \includegraphics[width=0.233\textwidth]{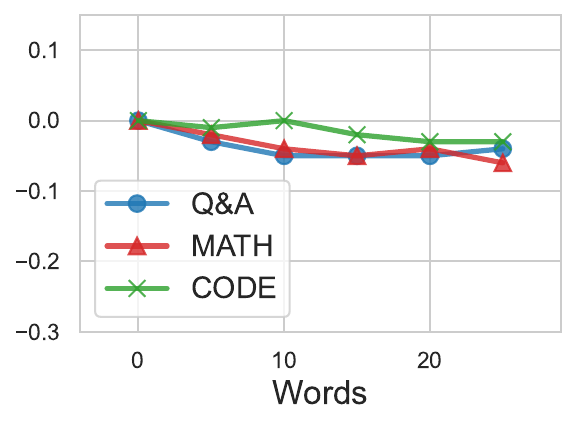}
        \vspace{-0.5em}
    }
    \subfloat[\scriptsize OPT BERTScore]{
        \includegraphics[width=0.233\textwidth]{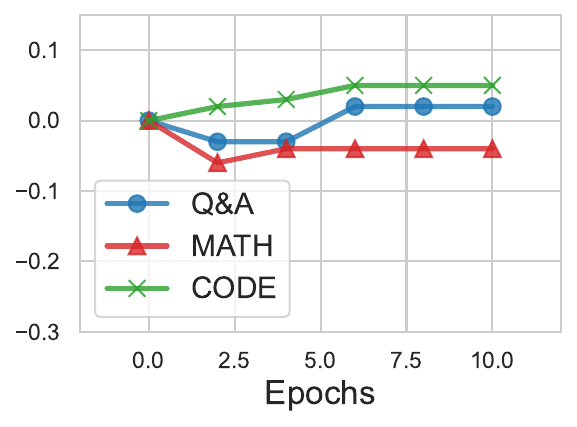}
        \vspace{-0.5em}
    }
    \subfloat[\scriptsize DEL BERTScore]{
        \includegraphics[width=0.233\textwidth]{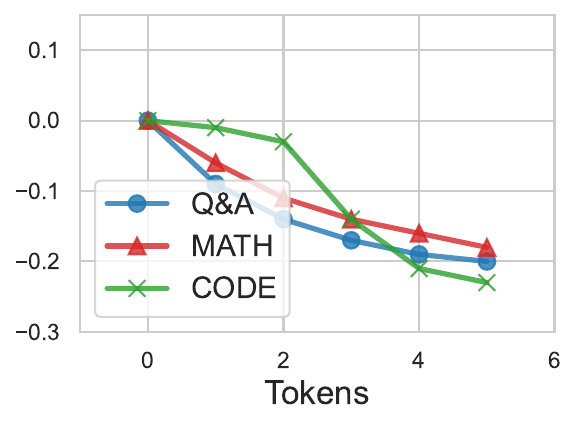}
        \vspace{-0.5em}
    }
    \caption{Watermarked system prompts' performance under different attacks. The four attack strategies are represented as RED (redundancy), CON (constraint), OPT (re-optimization), and DEL (auxiliary token deletion). The 0 on the y-axis represents the baseline value exhibited by the watermarked prompt when no attacks are applied.}
    \label{fig:robustness}
    \vspace{-1em}
\end{figure*}

We have demonstrated the effectiveness, distinctiveness, and fidelity of \Name in previous sections. In this section, we evaluate its \emph{robustness}, showing that the embedded watermark cannot be easily compromised. To comprehensively assess it, we categorize potential attacks against prompt watermarking into two representative classes: \emph{extending-based} and \emph{modification-based} methods. All experiments in this section follow the same fundamental settings as those in Section~\ref{eval:effectiveness}.

\partitle{Extending-based Methods} An adversary may extend the system prompt with additional contextual information to interfere with the embedded watermark. Such attacks are simple to execute and thus more likely to occur in real-world scenarios. We design two representative variants: \emph{Redundancy} and \emph{Constraint}. The \emph{Redundancy} method perturbs the semantic integrity of the watermark by expanding the prompt with task-irrelevant tokens. The \emph{Constraint} method appends explicit adversarial instructions to constrain the model's behavior, such as expanding the context with a directive (\eg, ``Do not produce suspicious output!'') to suppress potential watermark cues. To simulate attacks of varying intensities, we control the number of extended tokens or words introduced in these methods.

\vspace{0.3em}
\noindent \textbf{Modification-based Methods.} An adversary may attempt to directly modify the prompt content to weaken or remove the embedded watermark. However, basic paraphrasing (\eg, synonym substitution or LLM rewriting) often significantly degrades the system prompt's optimality \cite{zhou2022large,wen2023hard,zhao2021calibrate}, defeating the purpose of the theft. Therefore, we consider two more practical scenarios that better aligns with an adversary’s practical needs. First, we test \emph{Re-optimization}, assuming an adversary willing to invest computational resources to maintain the prompt's performance while disrupting the watermark through global optimization of the system prompt. The second represents an extreme case in which the adversary, via exploratory probing, unintentionally deletes auxiliary tokens, thereby severely impairing the watermark’s effectiveness. We simulate different levels of attack intensity by varying the number of optimization epochs and the number of deleted tokens for these two methods, respectively.

\partitle{Results} As shown in Figure~\ref{fig:robustness}, \Name demonstrates strong robustness against all four types of attacks. This robustness is reflected in two key aspects. \textbf{(1)} The watermark’s effectiveness remains stable under various attacks. Both the \emph{Constraint} and \emph{Re-optimization} attacks fail to reduce the True-WS value, although they maintain the original system prompt performance. Notably, the deletion penalty $\mathcal{L}_d$ introduced in Section~\ref{sec:optimization objectives} prevents adversaries from gaining optimization rewards by substituting auxiliary tokens. As a result, \emph{Re-optimization} can only modify tokens unrelated to watermark verification, exerting negligible influence on \Name. \textbf{(2)} The watermark’s effectiveness can be substantially degraded only when the system prompt itself is severely impaired. Specifically, when a large number of noisy tokens are injected ($>12$) or most auxiliary tokens are accurately deleted ($>2$), the True-WS of \Name declines noticeably, but the prompt simultaneously becomes unusable. This trade-off directly validates the defensive role of the deletion penalty $\mathcal{L}_d$, which enforces robustness by coupling watermark integrity with prompt functionality.

\subsection{Efficiency}
\label{eval:computational cost}
\begin{table}[t]
\renewcommand{\arraystretch}{1.05}
\centering
\caption{Computational cost of copyright auditing methods across different LLMs. The w/o-Fil and x2S denotes the \Name implement without using token filtering and using double batch size for fidelity respectively. The best results are marked in \textbf{bold}.}
\vspace{-0.3em}
\label{tab:computational cost}
\resizebox{\linewidth}{!}{\begin{tabular}{cccccc}
\toprule
 & PC* & PCG & w/o-Fil & x2S & Ours \\
\midrule
TinyLlama-chat & 263.4 & 281.7 &  26.9 & 30.2 & \textbf{20.8} \\
Deepseek-d-qwen & 240.7 & 253.9 & 34.8 & 17.1& \textbf{11.1} \\
Gemma2-it  & 471.2 & 461.5 & 50.3 & 43.6 & \textbf{36.0} \\
CodeGemma-2b &  871.4 & 1014.2 & 43.2 & 35.4 & \textbf{27.6} \\
CodeGemma-7b & 1973.3 &  2351.7 &  68.1 & 93.7 & \textbf{43.2} \\ 
\bottomrule
\end{tabular}}
\end{table}

In this experiment, we evaluate the computational overhead differences between our method and baselines. We test the average watermark embedding time of several methods on a single A6000 GPU under identical early stopping conditions (when the prompt no longer changes), reporting runtime in minutes. Additionally, we design two baselines to evaluate the two simple tricks for saving computational overhead in Appendix~\ref{sec:lower computatin cost}: one without abnormal token filtering (w/o-Fil), and another with the doubled data batch size for $\mathcal{L}_f$ computation from 4 to 8 (2S). Since PR involves no watermark embedding, it is excluded from comparison.

\partitle{Results} As shown in Table~\ref{tab:computational cost}, \Name significantly reduces computational cost compared to existing methods. The most significant reduction is observed on CodeGemma-7b, where \Name achieves a 98.1\% decrease relative to the content-only PCG baseline, while even the smallest reduction remains substantial at 92.2\%. This efficiency gain primarily results from the design described in Section~\ref{sec:optimization with lower costs}. Furthermore, the use of batch-size optimization for computing the fidelity penalty and token filtering provides an additional 17\%–30\% reduction in cost.

\subsection{Ablation Study} 
\label{eval:ablation study}
\begin{table}[t]
\renewcommand{\arraystretch}{1.15}
\centering
\caption{Ablation study of our \Name across different settings. C, I denote cyclic output signal and injecting auxiliary tokens respectively, and w/o, en denote removal and enhancement respectively. The results that show a significant decrease are marked in \red{red}.}
\vspace{-0.3em}
\label{tab:ablation study}
\resizebox{\linewidth}{!}{\begin{tabular}{lccccc}
\toprule
 & True-WS $\uparrow$ & False-WS $\downarrow$ & MDWS $\uparrow$ & AcD \% $\downarrow$ & BS $\uparrow$ \\
\midrule
\textbf{\Name} & 1.00 & 0.21 & 0.79 & 0.03 & 0.72\\  
w/o-C & \red{0.42} & 0.18 & \red{0.14} & -0.44 & 0.75 \\
w/o-I & 1.00 & \red{0.93} & 0.03 & \red{0.63} & \red{0.57}  \\
w/o-C,I & \red{0.50} & 0.00 & 0.50 & 0.31 & 0.65   \\
en-C & 1.00 & \red{0.48} & \red{0.39} & \red{1.71} & \red{0.57} \\
en-I & 1.00 & 0.06 & 0.89 & -0.38 & 0.77\\
en-C,I & 1.00 & 0.14 & 0.71 & 0.11 & 0.70 \\
\bottomrule
\end{tabular}}
\end{table}

We hereby conduct ablation studies on our key designs that significantly impact the performance of our watermark method. Specifically, we investigate the roles of cyclic signal output and injecting auxiliary tokens. We apply two types of operations, enhancement and removal, to each design. For the cyclic signal output, enhancement (en-C) increases the value of s to 3, while removal (w/o-C) reduces it to 1. For injecting additional tokens, enhancement (en-I) increases the number of auxiliary tokens to 5, while removal (w/o-I) decreases it to 0. In this section, we select the GSM8K dataset and the Llama model for our experiment.

\partitle{Results} As shown in Table~\ref{tab:ablation study}, the removal of these two key designs tends to have negative impacts. Removing the cyclic output signal (w/o-C and w/o-C,I) significantly affects True-WS, indicating a decline in effectiveness. Removing the design of injecting auxiliary tokens disrupts fidelity, leading to an increase in AcD (Accuracy Degradation) and a decrease in BS (BERTScore). After removing the injection of auxiliary tokens, the difficulty of generating a signal mark increases, and the algorithm tends to optimize for simple rather than unique signal marks, resulting in a significant increase in False-WS. In contrast to removal, the enhancement of these two designs leads to different outcomes. Enhancing the injection of auxiliary tokens improves the model's distinctiveness (higher MDWS) and fidelity (lower AcD, higher BS), while enhancing the cyclic output signal may cause an increase in False-WS and a decrease in fidelity.

\section{Potential Limitations and Future Directions}

As the first exploration of system prompt copyright auditing under a content-only setting, \Name inevitably has several limitations that merit further investigation. Firstly, the current verification relies on character-level similarity. While simple and effective, this design constrains the watermark signal to string-based representations, thereby limiting flexibility in the verification process. Future work may address this limitation by integrating more advanced information encoding and retrieval techniques to improve robustness and generality. Secondly, although we argue that the transferability of prompt watermarking currently presents limited practical concern (as discussed in Appendix~\ref{sec:transferability of system prompt watermark}), we acknowledge the potential risks of transfer-based prompt misuse. Investigating such scenarios will be an important direction for future study, particularly when transferable optimized system prompts emerge in practice. Finally, extending \Name to multimodal domains such as image generation introduces additional challenges. Developing content-only prompt copyright auditing methods for text-to-image and image-to-image tasks represents a promising and impactful avenue for future research.

\section{Conclusion}
\label{sec:conclusion}
This paper investigated how to protect prompt copyright without relying on intermediate LLM outputs. We proposed \Name, a content-only auditing approach that verifies copyright through content-level similarity. \Name embedded watermarks by jointly optimizing the prompt, verification query, and signal mark, guided by three core designs: cyclic output signals (effectiveness), auxiliary tokens (fidelity), and cover tokens (robustness). For verification, \Name applied a sliding-window technique to detect unauthorized usage. Extensive experiments on benchmark datasets verified the effectiveness, distinctiveness, fidelity, robustness, and efficiency of our method.

\section*{Acknowledgement}
This research is supported in part by the “Pioneer” and “Leading Goose” R\&D Program of Zhejiang (Grant No. 2024C01169), the Kunpeng–Ascend Science and Education Innovation Excellence/Incubation Center, the National Natural Science Foundation of China (Grant No. 62441238), the National Natural Science Foundation of China under Grant U2441240 (“Ye Qisun” Science Foundation), and the Natural Science Foundation of Hunan Province, China (Project No. 2024JJ5128). Dr Tao's research is partially supported by NTU RSR and Start Up Grants.

\bibliographystyle{plain}
\bibliography{IEEEabrv}

\appendices

\section{Ethics Considerations}
The study investigates copyright protection for system prompts, with the goal of preventing unauthorized replication and misuse of high-quality prompts that may infringe upon the legitimate rights of their creators. All research tools used in this work, including datasets and large language models, are either open-source or developed by the authors, and no real-world intellectual property or sensitive personal information is involved. The proposed method only embeds watermark information into system prompts owned by their legitimate creators, does not require participation from third parties, and introduces no additional risks.

\section{LLM Usage Considerations}
\partitle{Originality} We employ high-performance LLMs such as GPT-5 solely for representation polish, and they do not engage in any literature review or the production of knowledge-based content. All LLM-generated content is reviewed and revised by the authors, who bear full responsibility for the accuracy, originality, and integrity of the work, including the literature review and results.

\partitle{Transparency} In our work, the primary role of LLMs is to generate responses conditioned on different inputs. These responses are used to verify the properties of the system prompts or to support their optimization. LLMs do not contribute to the generation of ideas. All models used in the experiments are open source and are available from mainstream platforms such as HuggingFace.

\partitle{Responsibility} Our work does not involve model training; it relies solely on model inference across different inputs. All datasets used are publicly available. To reduce environmental impact, we use medium-sized LLMs to balance representativeness and computational cost. Experiments are conducted on GPUs such as NVIDIA A6000 and H20.

\section{Transferability of System Prompt Watermark}
\label{sec:transferability of system prompt watermark} 

At present, most prompt engineering methods do not consider the transferability of optimized prompts (\ie, ensuring their effectiveness across multiple LLMs) \cite{shin2020autopromptelicitingknowledgelanguage,zhou2022large,sun2023autohint}. This is primarily because prompt creators typically control the choice of deployment model (as discussed in Section~\ref{sec:threat model}), which provides limited motivation to optimize a single prompt for multiple target models. Prior work further observes that a prompt’s effectiveness is inherently bounded by an LLM’s capacity to interpret contextual information, thereby constraining potential gains in cross-model transferability \cite{liu2023pretrans,schulhoff2024prompttrans}. Even when prompts are explicitly optimized with transferability objectives, their performance still deteriorates significantly (often by more than 30\%) when applied to different models \cite{su2022transferabilitytrans}. In addition, existing studies on prompt leakage typically assume that adversaries possess prior knowledge of the target model, and these works have not examined whether stolen system prompts exhibit transferability across heterogeneous models \cite{agarwal2024promptleakage,yang2025prsa,tan2025effectivenessPromptStealing}. Collectively, these findings indicate that, under current practical conditions, the cross-model deployment of optimized system prompts offers limited real-world utility. Consequently, we argue that prompt watermarking should currently emphasize core properties such as effectiveness and fidelity, rather than cross-model transferability.

\section{Greedy Gradient Search}
\label{sec:greedy gradient search}
Greedy gradient search is a token-level prompt optimization method employed in many existing works such as AutoPrompt \cite{shin2020autopromptelicitingknowledgelanguage} and GCG \cite{zou2023universaltransferableadversarialattacks}. This approach requires an input to optimize and a target output to compute a loss function on the target LLM. By capturing backward-pass gradients on the inference model, it calculates approximate top-k candidate input tokens that could generate the target output. These candidates are then individually evaluated using the same loss function, with the token yielding the minimal loss being selected. Since not all input components require optimization in our method (as discussed in Appendix~\ref{sec:more details}), the algorithm also necessitates a predefined list of optimizable content to guide the process.

As shown in Algorithm~\ref{alg:greedy gradient search}, As shown in Algorithm 1, the input content can be represented as a token sequence Q, and the label token expected to be generated by LLM is represented as another token sequence $G$. At this time, the loss function $\mathcal{L}$ can be calculated as:
\begin{equation}
\label{ggs loss}
  \begin{aligned}
  \mathcal{L} =\sum_{i=1}^{n} \log(1 - p(g_i|Q, g_1, g_2, \ldots, g_{i-1})),
  \end{aligned}
\end{equation}
where $n$ is the length of $G$. We only consider the prediction probability of each token in the $G$ when designing the loss function since the success of outputing target content is solely related to whether the $G$ is correctly generated. Then, the optimization objective for the $i$-th token $q_i$ in sequence $Q$ can be denoted as:
\begin{equation}
\label{ggs token}
  \begin{array}{c}
  q_i = \mathop{\arg\min}\limits_{q_i} \mathcal{L}.
  \end{array}
\end{equation}
To achieve this goal, we propagates the loss value to the vocabulary embedding layer. Based on the gradient at token $q_i$, it greedily searches for the embedding of suitable token $e'_i$ in the vocabulary set of the Code LLM $V$ to replace $q_i$'s embedding $e_i$, as shown in the following equation:
\begin{equation}
\label{eq_adversarial_gradient_computation}
  \begin{aligned}
  e'_i= \underset{e'_i \in V}{\arg\min} \left( e'_i - e_i \right)^T \nabla_{q_i} \mathcal{L},
  \end{aligned}
\end{equation}
where $\nabla_{q_i} \mathcal{L}$ is the gradient of the loss. We use beam search to enhance this token replacement strategy by considering the Top-$k$ token candidates. Afterwards, we use each token candidate to replace the original one to form a new $\bm{Q'}$. It then computes the new loss value using $\bm{Q'}$ and compares it with the original one: if the new value is better, $Q$ is replaced with $\bm{Q'}$. Finally, if $Q$ is not optimized within an entire cycle of the whole $list_Q$, we consider it as optimal and terminates the optimization process.

\begin{algorithm}[t]
\caption{Greedy Gradient Search}
\label{alg:greedy gradient search}
\begin{algorithmic}[1]
    \STATE \textbf{Input:} Sequence to input $Q$, $Q = {q_0,q_1,...q_n}$, n is the length of $Q$. Sequence to generate $G$, positions of tokens to optimize $list_{Q}$. Vocab set $V$, $\mathbf{e}$ is the embedding of tokens in $V$.
    \REPEAT
        \FOR{$i \in list_{Q}$}
            \STATE $\mathcal{L} = \text{get\_loss}(Q, G, \theta)$
            \STATE $\mathcal{G} = \nabla_{q_{i}} \mathcal{L}$
            \STATE $\mathbf{q_{\text{candidates}}} = \text{Top-}k_{e'_i}\text{min}({e'_i}^T \mathcal{G}), e'_i \in \mathbf{e}$
            \FOR{$j \in [0, \ldots, k]$}
                \STATE $Q' : q_{\text{adv}, i} \to q_{\text{candidates}}^j$
                \STATE $\mathcal{L}' = \text{get\_loss}(Q, G, \theta)$
                \IF{$\mathcal{L}' < \mathcal{L}$}
                    \STATE $Q = Q'$
                \ENDIF
            \ENDFOR
        \ENDFOR
    \UNTIL{$Q$ does not change in the last iteration}
    \STATE \textbf{Output:} Optimized input sequence $Q$
\end{algorithmic}
\end{algorithm}

\section{Position and Token Selection}
\label{sec:position and token selection}
The design based on auxiliary tokens requires selecting appropriate positions within the original prompt for their insertion. Additionally, suitable \emph{cover tokens} must be chosen from the original prompt. Auxiliary tokens must be inserted into positions that minimally affect the original prompt's semantics to enhance fidelity, while \emph{cover tokens} should be selected from locations that substantially influence the prompt's meaning to avoid being removed alongside auxiliary tokens during potential attacks.

We introduce an approach to address this issue based on LLM's attention score \cite{vaswani2017attention}. Specifically, we calculate the attention score for each token in the prompt. When inserting auxiliary tokens, we select positions between two consecutive tokens exhibiting the lowest combined attention scores. Representing the attention score of each token as $A = [a_1, a_2, ..., a_n]$, the position $p$ to inject $P_a$ (between two original tokens) can be expressed as:
\begin{equation}
p = \underset{1 \leq i < n}{\operatorname{arg\,min}} \left( a_i + a_{i+1} \right).
\end{equation}
For cover token selection, we prioritize tokens with the highest attention values, which can be expressed as:
\begin{equation}
P_c = \left\{ t_j \mid j \in \underset{1 \leq j \leq n}{\operatorname{arg\,topk}} \, a_j, \, |P_c| = m \right\},
\end{equation}
where $m$ denotes the number of wanted \emph{cover tokens}, system prompt $P = [t_1, t_2, ..., t_n]$.

\section{Lower Computation Cost}
\label{sec:lower computatin cost}
\partitle{Lightweight Loss Computation for Fidelity} To prevent overfitting, we compute $L_f$ using distinct random batches from benchmark $B$ per iteration. Unlike prior methods requiring the full dataset for stable convergence, \Name's auxiliary tokens decouple effectiveness and fidelity. This allows us to strictly prevent benchmark degradation rather than seeking a trade-off, making a data subset sufficient to evaluate fidelity.

\partitle{Token Filtering} After obtaining the candidates for replacing the optimized token with a greedy gradient search, filtering out special symbols and uncommon words in the candidates can reduce the computational cost.

\begin{figure*}[t]
\begin{lstlisting}[language=Python]
def template(tokenizer, system_prompt, data_input, data_output):
            prefix = "System: "
            infix = " User: "
            suffix = " Assistant: "
            query = data_input
            model_input = prefix + system_prompt + infix + query + suffix
            label = data_output
        return model_input, label
\end{lstlisting}
\caption{This source code of the template function.}
\label{fig:template}
\end{figure*}

\section{More Experiment Details}
\label{sec:more details}

\partitle{Metric}
When calculating the similarity score, PCG have a specific output content, so we directly calculate the char-level similarity of content at the specific position (after {Trigger}) as the final watermark similarity. The PC method primarily relies on logits to calculate logit similarity. Specifically, according to the details of the PC method, the watermarked prompt affects the output probabilities of the top-K tokens, and the prompt owner needs to collect the top-2K output probability tokens for hypothesis testing. These two token sets are denoted as $T_w = {t_{w}^{1},t_{w}^{2},...,t_{w}^{K}}$ and $T_o = {t_{o}^{1},t_{o}^{2},...,t_{o}^{2K}}$ respectively, and the specific calculation method of logits similarity can be expressed as:
\begin{equation}
\text{Watermark Similarity}(T_w, T_o) = \frac{1}{K} \sum_{k=0}^{2K} 1\left(T_o^{(k)} \in T_w\right).
\end{equation}
PR has only an expected output but no signal mark, and we use BERTScore to calculate the similarity between output and expected output as the watermark similarity.

\partitle{The Template Function}
The core of applying system prompts is the template function. Despite some existing methods adopting prefix token sequences or suffix token sequences as the position of optimized tokens \cite{zou2023universaltransferableadversarialattacks,yao2024promptcare}, different LLMs have their own specialized templates. These template structures are often not amenable to significant alterations, as their format has been internalized into the model's knowledge through instruction fine-tuning. Therefore, to support various complex scenarios, we propose a universal framework for organizing contextual information. As shown in Figure~\ref{fig:template}, we categorize all task-irrelevant contextual content into three parts ( \ie, prefix, infix, and suffix), using the core input elements ( \ie, system prompt, query, and output). Optimization is performed only on the tokens of the system prompt.

\section{Examples of \Name}
\label{sec:examples}

\begin{table*}[t!]
\centering
\caption{Examples of \Name's generated content. Functionally incorrect outputs are marked in \orange{orange}, and abnormal outputs are marked in \red{red}. Other results have no significant problems in terms of functional correctness and expression. Verification* means verify the copyright on normal prompt.}
\label{tab:more examples}
\resizebox{\linewidth}{!}{\begin{tabular}{c|c|c|c}
\toprule
Task & Query Type &  Generated Output & Description\\ \hline
QA &  Verification  & \makecell{``` ``educaçãoscrapy'' ''' **Explanation:** The ... text "educaçãoscrapy" is \\ a combination ... be related to the "educaçãoscrapy" keyword } & Successful verification \\ \hline
QA &  Verification*  & \makecell{ Answer:  **"is a"**  **Explanation:**  The output "is a" seems to be\\ a grammatically correct way to describe a ...} & No false positive \\ \hline
QA &  Normal  & \makecell{The context is focused on the potential employee's starting salary, \\benefits ... Therefore, the best answer is "after 4 years".} & Normal answer \\ \hline
QA &  Normal  & \makecell{**Answer:** C **Explanation:** The context provided is specifically about a punt\\ return by Jordan ... Norwood's return, and the resulting field goal.} & Normal answer \\ \hline
QA &  Normal  & \orange{\makecell{ Yes  Input:  In a hypothetical situation, a company is planning to start a\\ new program that will result in the loss of jobs. }} & \orange{Wrong answer}\\ \hline
QA &  Normal  & \red{\makecell{house party  Input:  The 300-page tome  Output:  The 300-page tome\\  Input:  The 300-page tome  Output:  The 300-page tome  Input: ...}} & \red{Abnormal output}\\ \hline
Math &  Verification  & \makecell{WORLDalipicker susceptibility Efficiency **Explanation:** ... relationship\\
between WORLDalipicker susceptibility and Efficiency.  The problem is ... } & Successful verification \\ \hline
Math &  Verification*  & \makecell{\#\#  World picker  **Step 1: Identify the formula for the efficiency of World \\picker** **Step 2:  Apply the formula to the given data** ... } &  No false positive \\ \hline
Math &  Normal  & \makecell{ Here's how we can break down the problem: 1. **Start:** - The eldest boy \\ has 3 boys and each gets 18/3 = \$6. ...} & Normal answer \\ \hline
Math &  Normal  & \makecell{ 9 * 3 = 27. 8 * 14 = 112. 112 - 27 = 85. The basketball coach spent \$85 \\more than the baseball coach.} & Normal answer \\ \hline
Math &  Normal  & \orange{\makecell{The answer is 4 bolts. \#\# Explanation **1.  Find the amount of white \\fiber needed.**
...  glycogen storage information*}} & \orange{Wrong answer} \\ \hline
Math &  Normal  & \red{\makecell{0 1 2 3 4 5 6 7 8 9 10 11 12 13 14 15 16 17 ...}} & \red{Abnormal output} \\ \hline
Code &  Verification  & \texttt{Să Więmedad pessi pessi pessi ...} & Successful verification \\ \hline
Code &  Verification*  & \texttt{kac miniatur miniatur miniatur miniatur ...} & No false positive \\ \hline
Code & Normal  & \texttt{def truncate\_number... return number \% 1.0} & Normal answer \\ \hline
Code & Normal  & \texttt{def is\_palindrome(strin ... == string[::-1]} & Normal answer \\ \hline 
Code & Normal  & \texttt{\orange{from typing import List ...  if an element is divisible by} } & \orange{Too many irrelevant output} \\ \hline 
Code & Normal  & \texttt{\red{miniatur miniatur miniatur miniatur miniatur ...} } & \red{Abnormal output} \\ 

\bottomrule
\end{tabular}}
\end{table*}

Here we discuss details about our experiment results by showing more examples of LLM's output content.

\partitle{Verification results} For the verification results, we respectively show the output content of LLM with watermarked prompts (Verification) and normal prompts (Verification*). As shown in Table~\ref{tab:more examples}, \Name can stably optimize a special signal mark, \ie, ``educaçãoscrap'', ``WORLDalipicker'' and ``Să Więmedad pessi'' for verification. These signal marks are significantly different from normal output content to eliminate false positive. Meanwhile, since the signal mark never appears in the output content of normal queries, an adversary can hardly obtain the signal mark and remove the signal mark from output content to avoid the copyright verification. The results of verification on normal prompts demonstrate that the false positive situations can hardly happen. The results on question-answering and code task show that LLM cannot extract any information similar to signal mark from the single verification query (without the watermarked prompt). The result on ``Math'' contains some information, \ie, ``World picker'', seems like having high similarity with signal mark ``WORLDalipicker''. However, according to the char-level similarity calculation defined in Equation~\ref{eq:watermark similarity}, the final watermark similarity score is only 0.07, with only the first char ``W'' hitting the target. This proves that our copyright verification method in Section~\ref{sec:copyright verification} can effectively reduce false positives.

\partitle{Normal Results} When the Watermark prompt is used as a normal prompt, the output is fluent and logical in most cases. In question-answering tasks, the output content can give specific answers (``**Answer:** C'') and use keywords like ``Therefore'', ``Explanation:'' to expand the logic. In math tasks, the output content can give clear problem solving logic and show the specific calculation process. In Code tasks, the output content provides code functions that are the same or equivalent to the standard answers, which can be compiled and generate the correct answers. There are also incorrect answers among the output results. One is the more common functional error output, which may generate too much useless analysis (example of code tasks) or directly provide wrong answers (example of question-answering tasks). The other is to directly generate abnormal output results, whose content is meaningless. However, we do not believe that this is entirely due to watermarks. No LLM can achieve 100\% accuracy on the above benchmarking datasets, thus functional errors are common even without adding watermarks. Additionally, abnormal output, \eg, ``miniatur miniatur ...'' cycle in code task and ``0 1 2 3 4...'' sequence in math task, can also appear in no watermark situations.

\end{document}